\documentclass[twocolumn,english,aps,pra,superscriptaddress,address,showpacs,showacknowledgments]{revtex4-2}
\usepackage[T1]{fontenc}
\usepackage[active]{srcltx}
\usepackage{color}
\usepackage{amsmath}
\usepackage{amssymb}
\usepackage{stackrel}
\usepackage{graphicx}

\makeatletter
\usepackage{babel}

\makeatother

\usepackage{babel}
\begin{document}
\title{Single-photon scattering in giant-atom waveguide systems with chiral
coupling}
\author{Shu-Yu Li}
\affiliation{Fujian Key Laboratory of Quantum Information and Quantum Optics \&
Department of Physics, Fuzhou University, Fuzhou 350116, People's
Republic of China}
\author{Ze-Quan Zhang}
\affiliation{Fujian Key Laboratory of Quantum Information and Quantum Optics \&
Department of Physics, Fuzhou University, Fuzhou 350116, People's
Republic of China}
\author{Lei Du}
\affiliation{School of Physics and Center for Quantum Sciences, Northeast Normal
University, Changchun 130024, China}
\author{Yong Li}
\affiliation{Center for Theoretical Physics \& School of Physics and Optoelectronic
Engineering, Hainan University, Haikou 570228, China}
\author{Huaizhi Wu}
\affiliation{Fujian Key Laboratory of Quantum Information and Quantum Optics \&
Department of Physics, Fuzhou University, Fuzhou 350116, People's
Republic of China}
\begin{abstract}
We study single-photon scattering spectra of a giant atom chirally
coupled to a one-dimensional waveguide at multiple connection points,
and examine chirality induced effects in the scattering spectra. We
show that the transmission spectra typically possess an anti-Lorentzian
lineshape with a nonzero minimum, but by engineering the chirality
of the multi-point coupling, the transmission spectrum of an incident
photon can undergo a transition from complete transmission to total
reflection at multiple frequency ``windows'', where the width of
the anti-Lorentzian lineshape for each of the window can be flexibly
tuned at a fixed frequency detuning. Moreover, we show that a perfect
nonreciprocal photon scattering can be achieved due to the interplay
between internal atomic spontaneous emission and the chirally external
decay to the waveguide, in contrast to that induced by the non-Markovian
retardation effect. We also consider the non-Markovian retardation
effect on the scattering spectra, which allows for a photonic band
gap even with only two chiral coupling points. The giant-atom-waveguide
system with chiral coupling is a promising candidate for realizing
single-photon routers with multiple channels.
\end{abstract}
\maketitle

\section{INTRODUCTION}

Waveguide quantum electrodynamics (QED) \citep{1-1,AFK-chiral-23,Sheremet2023},
which studies the interaction between atoms (or other quantum emitters)
and free propagating photons in a one-dimensional (1D) waveguide,
has been experimentally demonstrated in many state-of-the-art architectures,
such as trapped natural or artificial atoms (including quantum dots
\citep{Lodahl2015,Uppu2021,Coles2016,JalaliMehrabad2020}, diamond
defects \citep{Sipahigil2016,ArjonaMartinez2022,pasini2023nonlinear,Pompili2021},
superconducting qubits \citep{Blais2021}, and single organic molecules
\citep{Toninelli2021}) coupled with optical fibers \citep{1-9,1-10,1-11},
photonic crystal waveguides \citep{1-12Goban2014,1-13,1-14}, or microwave
transmission lines (TLs) \citep{1-15,1-16Hoi2012,1-17Hoi2013,1-18Loo2013,1-19Liu2017}.
Typically, the (natural) atoms are orders of magnitude smaller than
optical (microwave) wavelengths of the continuous bosonic modes in
the 1D waveguide, therefore, they can be viewed as point-like emitters
to justify the dipole approximation. Waveguide QED systems with natural
``small'' atoms can potentially be used to construct quantum network
\citep{AFK_EIT-14Kimble2008,AFK_EIT-15Duan2010,AFK_EIT-16Wehner2018}
and simulate quantum many-body physics \citep{AFK_EIT-17Bello2023,AFK_EIT-18Wang2021,AFK_EIT-19Mahmoodian2020}.
On the other hand, extending the small atom platform to artificial
\textquoteleft giant\textquoteright{} atomic systems has attracted
significant recent attention (see the first review by Kockum \citep{AFK-FiveYear}),
in part because it represents a breakdown of the dipole approximation
where the scale of atoms becomes comparable to the wavelength of the
light they interact with. The artificial giant atoms have been well
designed and recently demonstrated with superconducting qubits coupled
to short-wavelength surface acoustic waves (SAWs) \citep{AFK_ArtificialAtom,AFK_Circuit,non-3GiantAcoustic,non-4non-exponentialDecay,experiment_EIT-SAW,YouJQ-experiment}
or a microwave-waveguide at multiple discrete points \citep{AFK_LambShift,AFK_experiment-Kannan2020,7Engineering}.
The multiple coupling points give rise to self-interference effects,
which are quite different from the conventional interference effects
among point-like small atoms (or resonators) coupled locally to a
common bath. The self-interference effects, which depend on both the
distances between coupling points and the photonic frequency, allows
to observe several unconventional phenomena, including frequency-dependent
Lamb shift and relaxation rate \citep{AFK_LambShift,LvXY}, decoherence-free
atomic states \citep{AFK_Decoherence,Carollo-decoherence,AFK_experiment-Kannan2020,Du_decoherence_arXiv,DL_PRR23},
non-Markovian decay dynamics \citep{non-1Beyond,non-2collisional,non-3GiantAcoustic,non-4non-exponentialDecay,non-5Oscillating},
and chiral light-matter interactions \citep{AFK-chiral,AFK_EIT-18Wang2021,2DuL_lambda,DuL-Synthetic,Chiral-Wang2022,LiaoJQ}.

Waveguide QED system with giant atoms has emerged as a new promising
platform for engineering transport of photons and single-photon routing
\citep{1DuL_lambda,2DuL_lambda,9DuL-,4JiaWZ-AnAtom,5Jia_TwoAtoms,24JiaWZ_NAtoms,WangZH_ATSplit,LiaoJQ_SinglePhoton,LiaoJQ,AFK_EIT_arXiv,AFK-chiral,LvXY}.
The system enables strong tunable atom-waveguide coupling and the
engineering of time delay, manifesting multiple-point interference
and non-Markovian retardation effects in the photon scattering spectra
\citep{2DuL_lambda,9DuL-,4JiaWZ-AnAtom,LiaoJQ_SinglePhoton,LiaoJQ}.
Herein, the photon scattering spectra can exhibit interesting features,
such as electromagnetically induced transparency \citep{5Jia_TwoAtoms,24JiaWZ_NAtoms,AFK_EIT_arXiv},
atomic decay induced nonreciprocity \citep{2DuL_lambda,9DuL-,LiaoJQ},
and photonic band gap \citep{4JiaWZ-AnAtom,24JiaWZ_NAtoms}, and these
features can be applied to probe collective radiance and topological
states \citep{5Jia_TwoAtoms,24JiaWZ_NAtoms,AFK-chiral} with a chain
of two-level giant atoms in both the Markovian and non-Markovian regimes
\citep{LvXY}. Moreover, an experimental setup with chiral interfaces
between giant atoms and waveguides has recently become a reality based
on technological progress, e.g., by coupling transmon qubits \citep{AFK-chiral-69,Joshi2023}
to meandering TLs with circulators \citep{AFK-chiral-46,AFK-chiral-47,AFK-chiral-48,AFK-chiral-49}.
In these setups, the coupling between waveguide modes and giant atoms
depends on the propagation direction of the light. Despite the chirality
of their coupling, one can still observe perfect collective radiance
and decoherence-free dark states \citep{AFK-chiral} inaccessible
with small atoms, and furthermore non-Markovianity induced nonreciprocity
\citep{2DuL_lambda,LiaoJQ} and photon frequency conversion {[}\citealp{2DuL_lambda}{]},
whereas the previous inspiring studies about single-photon routing
are limited to special cases, e.g., a (multilevel) giant atom with
two asymmetric coupling points \citep{1DuL_lambda,2DuL_lambda}. Thus,
it is unclear what can bring about by engineering the chirality with
more than two coupling points.

In this paper, we study single-photon scattering spectra and their
nonreciprocity with a two-level giant atom chirally coupled to a waveguide
at multiple points. We assume equally spaced coupling points, and
focus on chirality induced effects by considering three chiral coupling
regimes : (1) bidirectional even coupling (BEC), (2) unidirectional
un-even coupling (UUEC), and (3) bidirectional un-even coupling (BUEC),
respectively. By engineering the chirality of the couplings, we find
that an incident photon, which is fully transmitted in the BEC (UUEC)
regimes, can become totally reflected in the BUEC regime for a fixed
two-point propagating phase. This feature can not be observed in the
non-chiral setting simply by increasing the number of coupling points
$N$ \citep{4JiaWZ-AnAtom}. Moreover, the transmission spectra can
exhibit multiple ($N-1$) total reflection ``windows'' at fixed
frequencies, whose widths can be flexibly controlled. In comparison,
other chiral setups proposed for photon routing are involved with
a three-level giant atom \citep{2DuL_lambda} or giant-atom pairs
(i.e. giant molecules) \citep{LiaoJQ_SinglePhoton,LiaoJQ} and are
limited to the case of two coupling points, preventing it from multiple-window
photon routing. We examine the phases of the transmission and reflection
coefficients, and find that the giant atom imprints direction-dependent
phases on the incident photon uniquely for the chiral setup with uneven
couplings.

Furthermore, by taking atomic spontaneous decay into account, we find
that a perfect nonreciprocal photon scattering (with transmission
probability of unity in the forward direction and zero in the backward),
which is inaccessible in the giant-atom-waveguide system with uniformly
symmetric coupling \citep{4JiaWZ-AnAtom}, can now be realized by
engineering the chirality of the atom-waveguide interaction. This
is in contrast to other schemes concerning nonreciprocal photon scattering
\citep{2DuL_lambda,LiaoJQ}, where the nonreciprocity is typically
induced by non-Markovian retardation effects \citep{LiaoJQ} or by
synthetic gauge fields \citep{9DuL-}. We also consider the non-Markovian
retardation effect on the scattering spectra. We observe a photonic
band gap for the spectra in the intermediate non-Markovian regime
even with only two chiral coupling points, which normally arises in
a setup with large number of coupling point and with uniformly symmetric
coupling \citep{4JiaWZ-AnAtom}. The giant-atom-waveguide system with
chiral coupling provides a promising platform for realizing single-photon
routers with multiple frequency channels.

The paper is organized as follows: Section II introduces the theoretical
model, and calculates the transmission and reflection coefficients
of an incident photon by using a real-space scattering method, where
the giant atom's Lamb shifts and effective decay rates are derived
and discussed. Section III presents the scattering spectra for three
chiral coupling conditions (i.e. the BEC, the UUEC, and the BUEC conditions)
in the Markovian regime, and discusses intriguing features induced
by the chiral coupling. Section IV shows that nonreciprocal photon
scattering can be realized due to the cooperative effect of chiral
coupling and atomic spontaneous decay. Section V further discusses
the scattering spectra in the non-Markovian regime, with a conclusion
given by Section VI.

\section{MODEL AND METHOD}

\begin{figure}
\includegraphics[width=1\columnwidth]{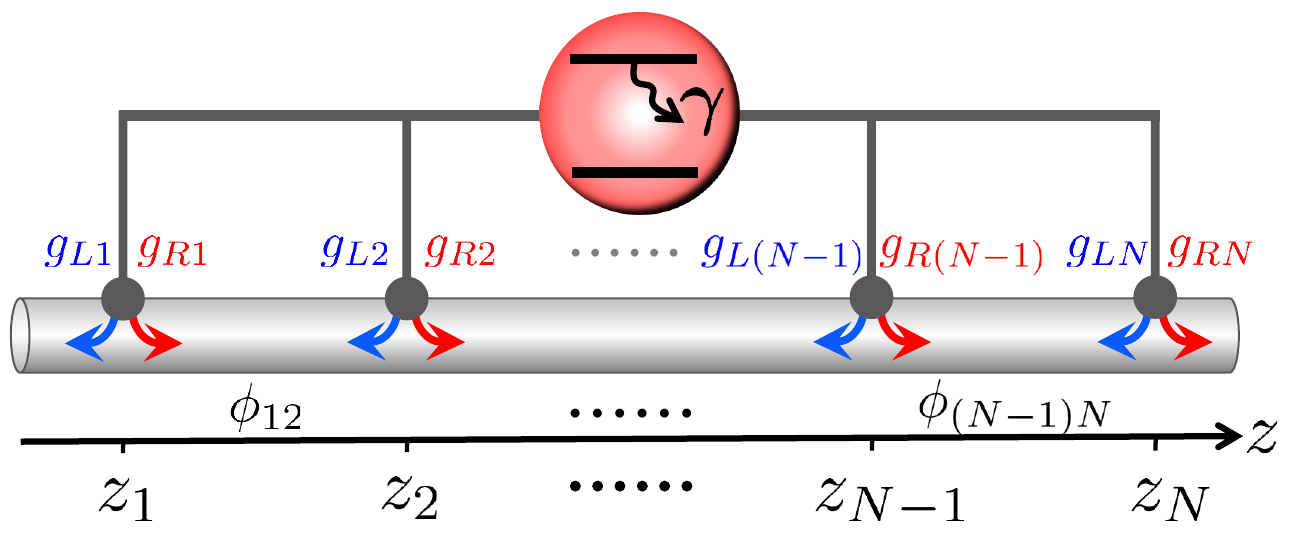}

\caption{\label{Fig.1Model}Schematics of a giant-atom system. A two-level
giant atom (with excited state $|e\rangle$ and ground state $|g\rangle$)
couples to the one-dimensional waveguide at multiple points (labeled
by the coordinates $z_{j}$), where $g_{Lj}$ ($g_{Rj}$) are coupling
strengths at the $j$th coupling point for left-going (right-going)
waveguide modes, $\phi_{ij}$ are distance-dependent phases for the
photon propagating between the $i$th and the $j$th coupling points,
and $\gamma$ is the spontaneous decay rate of the atom to the bath
environment.}
\end{figure}

As schematically shown in Fig. \ref{Fig.1Model}, the system consists
of a two-level giant atom chirally coupled to a 1D waveguide at $N$
discrete points (labeled $j$, coordinates $z_{j}$). The coupling
strengths between the giant atom and waveguide modes depend on the
propagation direction of the light. Under the rotating-wave approximation,
the total Hamiltonian of the system in real space can be written as
($\hbar=1$ hereafter) \citep{AGT_few-photon} 
\begin{equation}
H=H_{a}+H_{w}+H_{int},\label{eq:Hamiltonian}
\end{equation}
with 
\begin{align*}
H_{a}= & (\omega_{e}-i\gamma)|e\rangle\langle e|,\\
H_{w}= & \int_{-\infty}^{+\infty}dza_{L}^{\dagger}\left(z\right)\left(\omega_{0}+iv_{g}\frac{\partial}{\partial z}\right)a_{L}\left(z\right)\\
 & +a_{R}^{\dagger}\left(z\right)\left(\omega_{0}-iv_{g}\frac{\partial}{\partial z}\right)a_{R}\left(z\right),\\
H_{int}= & \int_{-\infty}^{+\infty}dz\stackrel[j=1]{N}{\sum}\delta(z-z_{j})\sqrt{v_{g}}\left[g_{Lj}a_{L}^{\dagger}\left(z\right)e^{ik_{0}z}\right.\\
 & +\left.g_{Rj}a_{R}^{\dagger}\left(z\right)e^{-ik_{0}z}\right]|g\rangle\langle e|+\text{H.c.},
\end{align*}
where $H_{a}$ is the bare atomic Hamiltonian, with $\omega_{e}$
being the transition frequency between the ground state $|g\rangle$
and the excited state $|e\rangle$, and $\gamma$ being the spontaneous
emission rate induced by the non-waveguide modes in the environment.
$H_{w}$ is the bare waveguide Hamiltonian, with $a_{L}^{\dagger}\left(z\right)$
$[a_{R}^{\dagger}\left(z\right)]$ and $a_{L}(z)$ {[}$a_{R}(z)${]}
the creation and annihilation operators of the left-propagating (right-propagating)
modes at position $z$. $\omega_{0}$ is the central frequency around
which a linear dispersion relation under consideration is given by
$\omega(k)=\omega_{0}+\left(k-k_{0}\right)v_{g}$ with $k$ the wave
vector of the incident photon, $k_{0}$ the wave vector corresponding
to $\omega_{0}$, and $v_{g}$ the group velocity in the vicinity
of $\omega_{0}$ \citep{Shen2009}. $H_{int}$ is the interaction
Hamiltonian with $g_{Lj}$ and $g_{Rj}$ the renormalized coupling
strengths for the atom interacting with a left-going and right-going
photon at the position $z=z_{j}$, respectively, where in real space,
the giant atom behaves as a potential $\delta\left(z-z_{j}\right)$
at each coupling point. Note that the atom-waveguide coupling at multiple
points enable us to observe system dynamics in both the Markovian
and non-Markovian regimes \citep{AFK-chiral,LiaoJQ}, which strongly
depends on the accumulated phases $\phi_{ij}=k|z_{i}-z_{j}|$ {[}or
$\phi_{ij}=(\omega-\omega_{0})\tau_{ij}+k_{0}|z_{i}-z_{j}|${]} of
photons propagating between any two of the $N$ coupling points with
$\tau_{ij}=\left|z_{i}-z_{j}\right|/v_{g}$.

We consider the single-photon scattering problem by engineering chirality
of the atom-waveguide couplings, where the system is constrained to
the single excitation subspace, and then the eigenstate of Hamiltonian
(\ref{eq:Hamiltonian}) is given by 
\begin{eqnarray}
\left|\psi\right\rangle  & = & \int_{-\infty}^{+\infty}dz\left[c_{gL}\left(z\right)a_{L}^{\dagger}\left(z\right)+c_{gR}\left(z\right)a_{R}^{\dagger}\left(z\right)\right]|g,0\rangle\nonumber \\
 &  & +c_{e0}|e,0\rangle,\label{eq:eigenstate}
\end{eqnarray}
where $c_{gL}\left(z\right)$ {[}$c_{gR}\left(z\right)${]} is the
probability amplitude of the states $a_{L}^{\dagger}\left(z\right)|g,0\rangle$
{[}$a_{R}^{\dagger}\left(z\right)|g,0\rangle${]}, describing a left-propagating
(right-propagating) photon at position $z$ and the atom in $|g\rangle$,
and $c_{e0}$ is the probability amplitude of the atom in the excited
state $|e\rangle$, finding no photon in the waveguide.

Solving the stationary Schrödinger equation $H|\psi\rangle=E|\psi\rangle$
with an appropriate ansatz for the probability amplitudes $c_{gL}\left(z\right)$
and $c_{gR}\left(z\right)$ (see Appendix A), we obtain the transmission
and reflection coefficients for a left-incident photon 
\begin{align}
t_{N} & =\frac{\left(\Delta-\Delta_{ls}\right)+i\left(\gamma+\Gamma_{L}-\Gamma_{R}\right)}{\left(\Delta-\Delta_{ls}\right)+i\left(\gamma+\Gamma_{L}+\Gamma_{R}\right)},\nonumber \\
r_{1} & =\frac{-i\Gamma_{LR}}{\left(\Delta-\Delta_{ls}\right)+i\left(\gamma+\Gamma_{L}+\Gamma_{R}\right)},\label{eq:TR_left}
\end{align}
and for a right-incident photon: 
\begin{align}
\tilde{t}_{1} & =\frac{\left(\Delta-\Delta_{ls}\right)+i\left(\gamma-\Gamma_{L}+\Gamma_{R}\right)}{\left(\Delta-\Delta_{ls}\right)+i\left(\gamma+\Gamma_{L}+\Gamma_{R}\right)},\nonumber \\
\tilde{r}_{N} & =\frac{i\Gamma_{LR}^{*}}{\left(\Delta-\Delta_{ls}\right)+i\left(\gamma+\Gamma_{L}+\Gamma_{R}\right)},\label{eq:TR_right}
\end{align}
where $\Delta=\omega(k)-\omega_{e}$ is the detuning between the incident
photons and the atomic transition $|g\rangle\leftrightarrow|e\rangle$;
$\Delta_{ls}\equiv\Delta_{L}+\Delta_{R}$ is the overall Lamb shift
contributed by both the left- and right-propagating directions waveguide
modes from interference between connection points, with 
\begin{equation}
\Delta_{L}=\frac{1}{2}\stackrel[i,j]{N}{\sum}g_{Li}g_{Lj}\text{sin}\phi_{ij},\text{ }\Delta_{R}=\frac{1}{2}\stackrel[i,j]{N}{\sum}g_{Ri}g_{Rj}\text{sin}\phi_{ij},\label{eq:xLambShift}
\end{equation}
correspondingly; $\Gamma_{L}$ and $\Gamma_{R}$ are direction-dependent
relaxation rates given by \citep{4JiaWZ-AnAtom,AFK_LambShift} 
\begin{equation}
\Gamma_{L}=\frac{1}{2}\stackrel[\{i,j\}=1]{N}{\sum}g_{Li}g_{Lj}\text{cos}\phi_{ij},\label{eq:xDecayRate}
\end{equation}
\begin{equation}
\Gamma_{R}=\frac{1}{2}\stackrel[\{i,j\}=1]{N}{\sum}g_{Ri}g_{Rj}\text{cos}\phi_{ij},\label{eq:yDecayRate}
\end{equation}
and 
\begin{equation}
\Gamma_{LR}=\left(\stackrel[i=1]{N}{\sum}e^{ikz_{i}}g_{Li}\right)\left(\stackrel[j=1]{N}{\sum}e^{ikz_{j}}g_{Rj}\right).\label{eq:xyDecayRate}
\end{equation}
It follows that the transmission and reflection probabilities are
defined by $\mathcal{T}_{L}=\left|t_{N}\right|^{2}$ ($\mathcal{T}_{R}=\left|\tilde{t}_{1}\right|^{2}$)
and $\mathcal{R}_{L}=\left|r_{1}\right|^{2}$ ($\mathcal{R}_{R}=\left|\tilde{r}_{N}\right|^{2}$),
respectively. Note that the (chiral) atom-photon interaction is imprinted
on the phases of the transmission and reflection coefficients, which
depend on the Lamb shift $\Delta_{ls}$ and the direction-dependent
relaxation rates $\Gamma_{L}$ and $\Gamma_{R}$. In particular, for
$\mathcal{T}_{L(R)}=1$ or $\mathcal{R}_{L(R)}=1$, the phase imprinted
on the scattering photon can depend on the incident direction and
the chirality of couplings, see further discussion later.

Revisiting the small-atom limit \citep{AFK_LambShift}, where the
atom interacts with the left-going (right-going) waveguide modes by
a single connection point (e.g., at position $z_{j}$), the atomic
relaxation rates into the continuum modes can be simply derived by
Fermi's golden rule and are given by $g_{Lj}^{2}/2$ and $g_{Rj}^{2}/2$,
and the atomic ground state is not shifted by the atom-waveguide coupling
(i.e., $\Delta_{L(R)}=0$) \citep{AFK_LambShift}. In the case of
a giant-atom with $N$ connection points, when the couplings with
left-going and right-going photons are uniformly symmetric, i.e.,
$g_{Li}=g_{Lj}=g_{Ri}=g_{Rj}$ (for all $i\neq j$), one finds $\Delta_{L}=\Delta_{R}$
and $\Gamma_{L}=\Gamma_{R}$ regardless of the details of phases $\phi_{ij}$,
then the transmission and reflection probabilities read 
\begin{equation}
\mathcal{T}_{L}=\mathcal{T}_{R}=\frac{\left(\Delta-2\Delta_{L}\right)^{2}+\gamma^{2}}{\left(\Delta-2\Delta_{L}\right)^{2}+\left(\gamma+2\Gamma_{L}\right)^{2}},\label{eq:non-chiral-T}
\end{equation}
\begin{equation}
\mathcal{R}_{L}=\mathcal{R}_{R}=\frac{4\Gamma_{L}^{2}}{\left(\Delta-2\Delta_{L}\right)^{2}+\left(\gamma+2\Gamma_{L}\right)^{2}}.\label{eq:non-chiral-R}
\end{equation}
For $\gamma=0$, the reflection probabilities have the standard Lorentzian
lineshapes centered at $\Delta=2\Delta_{L}$ with the full-width at
half-maximum (FWHM) $4\Gamma_{L}$ and the transmission probabilities
have the anti-Lorentzian lineshapes. There exists a singular point
with $\Gamma_{L}=0$ for which the giant atom decouples from the waveguide,
leading to $\mathcal{T}_{L}=\mathcal{T}_{R}=1$. Moreover, Eqs. (\ref{eq:non-chiral-T})
and (\ref{eq:non-chiral-R}) imply that nonreciprocal photon scattering
can not happen whether $\gamma=0$ or $\gamma\neq0$.

To clearly see chirality induced features, we now consider the $N$
coupling points being equally spaced and firstly focus on the effect
of even or uneven chiral coupling in the Markovian regime, where the
propagating time of photons travel between the leftmost and the rightmost
coupling points $\tau_{1N}=\left(z_{N}-z_{1}\right)/v_{g}$ is short
compared to the characteristic relaxation time $\sim\tilde{\Gamma}^{-1}$
of the giant atom with $\tilde{\Gamma}=[(\sum_{i=1}^{N}g_{Li})^{2}+(\sum_{j=1}^{N}g_{Rj})^{2}]/2+\gamma$,
corresponding to the decay rate for $\tau_{1N}\rightarrow0$. However,
when $\tau_{1N}\sim1/\tilde{\Gamma}$, the giant atom-waveguide interaction
will enter the non-Markovian regime \citep{non-3GiantAcoustic,AFK_LambShift,4JiaWZ-AnAtom,AFK-FiveYear},
i.e., the time evolution of the system can depend on what the system
state was at an earlier time, and will be discussed later in Sec.
V. It should be emphasized that the transmission and reflection coefficients
in Eqs. (\ref{eq:TR_left})-(\ref{eq:TR_right}) are valid in both
the Markovian and the non-Markovian regimes. Moreover, given that
the bandwidth $\Delta$ is of the order of $\tilde{\Gamma}$, we can
examine the Markovian physics based on Eqs. (\ref{eq:xLambShift})-(\ref{eq:yDecayRate})
under the condition of $\tilde{\Gamma}\tau_{1N}\sim|\Delta|\tau_{1N}\ll1$,
and neglect the contribution of $|\Delta|\tau_{1N}$ to the $\phi_{ij}$
dependent effect. In the next section, we first replace the phases
$\phi_{ij}$ by $\tilde{\phi}_{ij}=(\omega_{e}-\omega_{0})\tau_{ij}+k_{0}|z_{i}-z_{j}|$,
and the transmission and reflection coefficients are rephrased in
terms of the phase difference between adjacent coupling points $\tilde{\phi}_{12}$,
i.e., $\phi_{ij}=|j-i|(\Delta\tau_{12}+\tilde{\phi}_{12})\rightarrow\tilde{\phi}_{ij}=\left|j-i\right|\tilde{\phi}_{12}$.
We take $\tilde{\phi}_{12}$ modulo $2\pi$ {[}denoted by mod$(\tilde{\phi}_{12},2\pi)$
in the following{]} into account for its effectiveness, and study
photon scattering under different chiral coupling regimes.

\begin{figure}
\begin{centering}
\includegraphics[width=1\columnwidth]{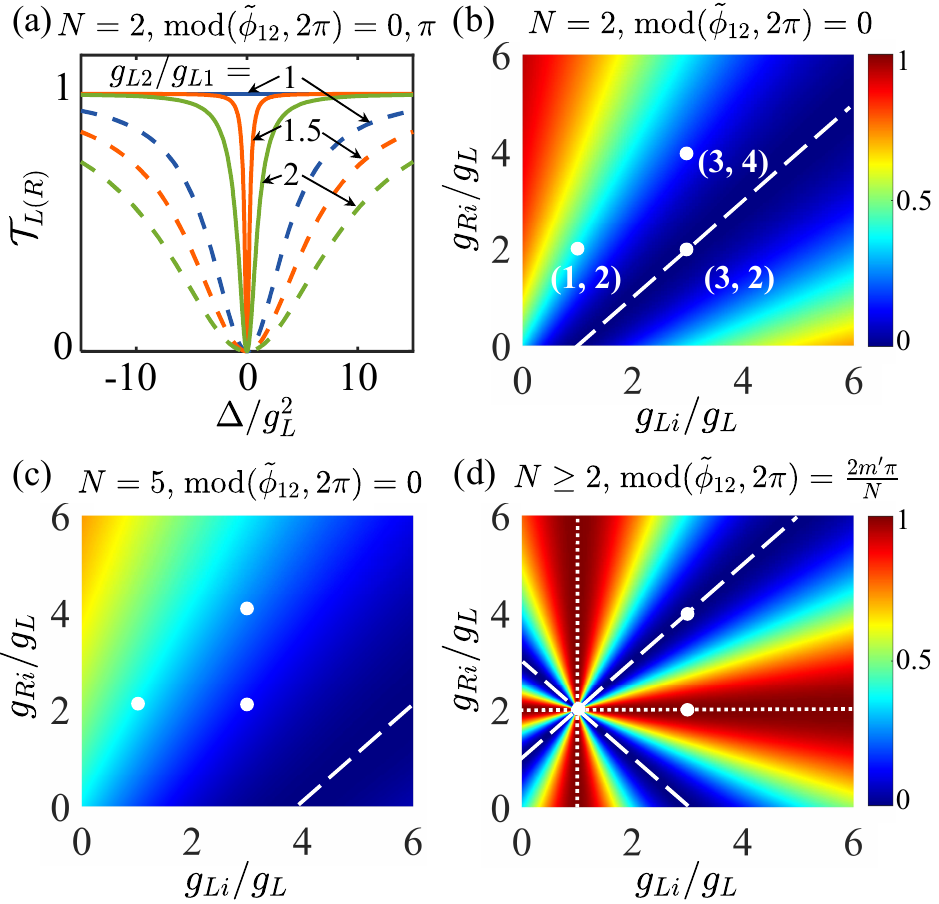}
\par\end{centering}
\caption{\label{fig2_Txy}Transmission probabilities $\mathcal{T}_{L(R)}$
in the Markovian regime with chiral coupling. (a) $\mathcal{T}_{L(R)}$
given by Eq. (\ref{eq:Ta_N2_Uneven2-1}) and Eq. (\ref{eq:Tb_N2_Uneven2})
as functions of $\Delta$ by considering $g_{L2}/g_{L1}=\{1,1.5,2\}$
with $N=2$ coupling points. The dashed lines correspond to the single-photon
transmission spectrum under the condition of $g_{L1}+g_{L2}=g_{R1}+g_{R2}$
and $\text{mod}(\tilde{\phi}_{12},2\pi)=0$, and the solid lines correspond
to $|g_{L1}-g_{L2}|=|g_{R1}-g_{R2}|$ and $\text{mod}(\tilde{\phi}_{12},2\pi)=\pi$.
For $\text{mod}(\tilde{\phi}_{12},2\pi)=\pi$, $\mathcal{T}_{L(R)}$
remains unity under the even coupling condition $g_{L1}=g_{L2}$,
while for $g_{L1}\protect\neq g_{L2}$, $\mathcal{T}_{L(R)}$ possess
the anti-Lorentzian lineshape of the width $2(g_{L1}-g_{L2})^{2}$
and undergo a sudden jump from unity to zero at resonance $\Delta=0$.
(b)-(d) $\mathcal{T}_{L(R)}$ versus the coupling strengths $g_{Li}/g_{L}$
and $g_{Ri}/g_{L}$ at the $i$th coupling point with (b) $N=2$,
$\text{mod}(\tilde{\phi}_{12},2\pi)=0$, $\Delta=\Delta_{ls}=0$,
(c) $N=5$, $\text{mod}(\tilde{\phi}_{12},2\pi)=0$, $\Delta=\Delta_{ls}=0$,
and (d) $N\protect\geq2$, $\text{mod}(\tilde{\phi}_{12},2\pi)=2m^{\text{\ensuremath{\prime}}}\pi/N$,
$\Delta=\Delta_{ls}$, respectively. Here, we consider the $N$ coupling
points are equally spaced, and set $(g_{Lj},g_{Rj})/g_{L}=(1,2)$
for $j\protect\neq i$. The white dotted lines indicate
$\mathcal{T}_{L(R)}=1$ under the condition of $g_{Li}=g_{L}$ or
$g_{Ri}=g_{R}$, i.e. the UUEC regime. The white dashed lines indicate
$\mathcal{T}_{L(R)}=0$ under the condition of $\Gamma_{L}=\Gamma_{R}$
corresponding to the BUEC regime, except the intersection (BEC) point
of the lines. The white dots in (b)-(d) indicate the set of coupling
strengths $(g_{Li},g_{Ri})/g_{L}=\{(1,2),(3,2),(3,4)\}$, with respect
to the BEC, UUEC, and BUEC regimes discussed later in Fig. \ref{Fig:T.vs.Delta_phi}.}
\end{figure}

\section{photon scattering with chiral coupling: The Markovian regime}

By considering the chiral coupling with $\gamma=0$ and assuming $g_{Li}$
($g_{Ri}$) being positive real values, one typically has $\Gamma_{L}\neq\Gamma_{R}$,
i.e., $\frac{1}{2}\sum_{\{i,j\}=1}^{N}(g_{Li}g_{Lj}-g_{Ri}g_{Rj})\text{cos}\phi_{ij}\neq0$.
The chirality can be divided into three different regimes: (1) bidirectional
even coupling (BEC) $g_{Li}=g_{Lj},g_{Ri}=g_{Rj},$ but $g_{Li}\neq g_{Ri}$;
(2) unidirectional un-even coupling (UUEC) $g_{Li}=g_{Lj},g_{Ri}\neq g_{Rj}$,
or $g_{Li}\neq g_{Lj},g_{Ri}=g_{Rj}$; (3) bidirectional un-even coupling
(BUEC) $g_{Li}\neq g_{Lj},g_{Ri}\neq g_{Rj}$. Remarkably, we find
that $\Gamma_{L}=\Gamma_{R}$ can be achieved by appropriately tuning
the uneven coupling strengths distributed at different coupling points.
As a result, it offers a flexible way to control the transmission
and reflection of a single incident photon, and allows for photon
routing at multiple fixed frequencies corresponding to $\Delta=\Delta_{ls}$.
This will be discussed in detail later.

\subsection{Bidirectional even coupling}

We first consider the BEC regime for a giant atom with $N\geq2$ coupling
points. The coupling strengths of the $N$ coupling points are identical
for the same propagation direction, and are set to $g_{Li}=g_{L}$
and $g_{Ri}=g_{R}$. Due to the multiple-point interference effect,
the overall Lamb shift takes the form 
\begin{equation}
\Delta_{ls}=\frac{1}{2}\left(g_{L}^{2}+y_{R}^{2}\right)\frac{N\text{sin}\tilde{\phi}_{12}-\text{sin}N\tilde{\phi}_{12}}{1-\text{cos}\tilde{\phi}_{12}},
\end{equation}
and the sum (difference) of the two effective decay rates are given
by 
\begin{eqnarray}
\Gamma_{L}\pm\Gamma_{R} & = & \frac{1}{2}\left(g_{L}^{2}\pm g_{R}^{2}\right)\frac{\text{sin}^{2}\left(\frac{1}{2}N\tilde{\phi}_{12}\right)}{\text{sin}^{2}\left(\frac{1}{2}\tilde{\phi}_{12}\right)},
\end{eqnarray}
which show that the photon interferes with itself multiple times independently
for the distinct (left and right) propagation directions.

(1) When the two-point propagating phase satisfies $\text{mod}(\tilde{\phi}_{12},2\pi)=0$,
the Lamb shift vanishes regardless of the chiral couplings, but the
effective decay rates are $N^{2}$-enhanced and are given by
\begin{equation}
\Gamma_{L}\pm\Gamma_{R}=\frac{1}{2}N^{2}\left(g_{L}^{2}\pm g_{R}^{2}\right).
\end{equation}
The transmission probabilities then read 
\begin{eqnarray}
\mathcal{T}_{L}=\mathcal{T}_{R} & = & \frac{4\Delta^{2}+N^{4}\left(g_{L}^{2}-g_{R}^{2}\right)^{2}}{4\Delta^{2}+N^{4}\left(g_{L}^{2}+g_{R}^{2}\right)^{2}}.
\end{eqnarray}
In contrast to the case with uniformly symmetric coupling ($g_{L}=g_{R}$),
the chirality of couplings leads to a nonvanishing transmission $\mathcal{T}_{L(R)}\neq0$
at resonance $\Delta=0$, i.e., 
\begin{equation}
\mathcal{T}_{L(R)}=\left(\frac{\Gamma_{L}-\Gamma_{R}}{\Gamma_{L}+\Gamma_{R}}\right)^{2}=\left(\frac{g_{L}^{2}-g_{R}^{2}}{g_{L}^{2}+g_{R}^{2}}\right)^{2},
\end{equation}
corresponding to $\mathcal{T}_{L(R)}$ in the large $N\rightarrow\infty$
limit.

(2) For $\text{mod}(\tilde{\phi}_{12},2\pi)=2m^{\text{\ensuremath{\prime}}}\pi/N$
($m^{\text{\ensuremath{\prime}}}=1,2,...,N-1$), the Lamb shift can
be nonvanishing and is given by 
\begin{equation}
\Delta_{ls}=\frac{N}{2}\left(g_{L}^{2}+g_{R}^{2}\right)\text{cot}\left(\frac{m^{\text{\ensuremath{\prime}}}\pi}{N}\right),\label{eq:Lamb2_BEC}
\end{equation}
but the effective decay rates vanish ($\Gamma_{L}=\Gamma_{R}=0$)
due to the destructive interference effect among the coupling points.
Then, a photon incident from the left or the right will be completely
transmitted, independent of the frequency detuning between the incident
photon and the atomic transition. This is analogous to the case with
uniformly symmetric coupling, except that $\Delta_{L}\neq\Delta_{R}$.

\subsection{Unidirectional un-even and bidirectional un-even coupling}

The results for the BEC regime can be understood intuitively, but
those for the UUEC and the BUEC regime are not self-evident. To obtain
instructive insight, we first consider the simplest model with only
$N=2$ coupling points in the Markovian limit, where the Lamb shifts
are $\Delta_{L}=\Delta_{R}=0$ both for $\text{mod}(\tilde{\phi}_{12},2\pi)=0$
and $\text{mod}(\tilde{\phi}_{12},2\pi)=2m^{\text{\ensuremath{\prime}}}\pi/N$,
but the sum and difference of the two effective decay rates $\Gamma_{L}\pm\Gamma_{R}$
strongly depend on the chirality of the coupling strengths. For $\text{mod}(\tilde{\phi}_{12},2\pi)=0$,
the effective decay rates are simply given by
\begin{equation}
\Gamma_{L}\pm\Gamma_{R}=\frac{(g_{L1}+g_{L2})^{2}\pm(g_{R1}+g_{R2})^{2}}{2},
\end{equation}
and correspondingly the transmission probabilities are 
\begin{eqnarray}
\mathcal{T}_{L(R)} & = & \frac{4\Delta^{2}+\left[\left(g_{L1}+g_{L2}\right)^{2}-\left(g_{R1}+g_{R2}\right)^{2}\right]^{2}}{4\Delta^{2}+\left[\left(g_{L1}+g_{L2}\right)^{2}+\left(g_{R1}+g_{R2}\right)^{2}\right]^{2}};\label{eq:Ta_N2_Uneven1}
\end{eqnarray}
for $\text{mod}(\tilde{\phi}_{12},2\pi)=\pi$, the effective decay
rates are alternatively given by 
\begin{eqnarray}
\Gamma_{L}\pm\Gamma_{R} & = & \frac{(g_{L1}-g_{L2})^{2}\pm(g_{R1}-g_{R2})^{2}}{2},\label{eq:chiral_N2-1}
\end{eqnarray}
and the transmission probabilities read 
\begin{equation}
\mathcal{T}_{L(R)}=\frac{4\Delta^{2}+\left[\left(g_{L1}-g_{L2}\right)^{2}-\left(g_{R1}-g_{R2}\right)^{2}\right]^{2}}{4\Delta^{2}+\left[\left(g_{L1}-g_{L2}\right)^{2}+\left(g_{R1}-g_{R2}\right)^{2}\right]^{2}}.\label{eq:Tb_N2_uneven0}
\end{equation}
Note that in the UUEC regime {[}i.e., $g_{L1}=g_{L2}$ and $g_{R1}\neq g_{R2}$
(or $g_{R1}=g_{R2}$ and $g_{L1}\neq g_{L2}$){]}, if the coupling
strengths satisfy $g_{L1}+g_{L2}=g_{R1}+g_{R2}$, Eq. (\ref{eq:Ta_N2_Uneven1})
reduces to 
\begin{equation}
\mathcal{T}_{L(R)}=\frac{\Delta^{2}}{\Delta^{2}+\left(g_{L1}+g_{L2}\right)^{4}},\label{eq:Ta_N2_Uneven2-1}
\end{equation}
which possesses the anti-Lorentzian line shape centered at $\Delta=0$
with the FWHM $2(g_{L1}+g_{L2})^{2}$ and the minimum $\mathcal{T}_{L(R)}(\Delta=0)=0$,
in contrast to $0<\mathcal{T}_{L(R)}<1$ of the BEC regime, see Fig.
\ref{fig2_Txy}(a); on the other hand, the transmission probabilities
{[}Eq. (\ref{eq:Tb_N2_uneven0}){]} for $\text{mod}(\tilde{\phi}_{12},2\pi)=\pi$
are constant unity (i.e., $\mathcal{T}_{L(R)}\equiv1$) regardless
of the specific value of the detuning $\Delta$, exhibiting the same
feature as in the BEC regime. In the BUEC regime, the overall profile
of $\mathcal{T}_{L(R)}(\Delta)$ for $\text{mod}(\tilde{\phi}_{12},2\pi)=0$
is below unity and $\mathcal{T}_{L(R)}(\Delta=0)=0$ again for $g_{L1}+g_{L2}=g_{R1}+g_{R2}$;
while for $\text{mod}(\tilde{\phi}_{12},2\pi)=\pi$, it is remarkable
that, under the condition of $|g_{L1}-g_{L2}|=|g_{R1}-g_{R2}|$ (corresponding
to $\Gamma_{L}=\Gamma_{R}\neq0$), $\mathcal{T}_{L(R)}(\Delta)$ {[}Eq.
(\ref{eq:Tb_N2_uneven0}){]} reduce to 
\begin{equation}
\mathcal{T}_{L(R)}=\frac{\Delta^{2}}{\Delta^{2}+\left(g_{L1}-g_{L2}\right)^{4}},\label{eq:Tb_N2_Uneven2}
\end{equation}
which have the minimum $\mathcal{T}_{L(R)}(\Delta=0)=0$ at resonance,
and possess the FWHM $2(g_{L1}-g_{L2})^{2}$ that can be infinitely
narrow for $g_{L2}/g_{L1}\rightarrow1$, in contrast to $\mathcal{T}_{L(R)}(\Delta)\equiv1$
for both the BEC and the UUEC regime. In other words, an incident
photon can be totally reflected or fully transmitted for the \textquotedblleft destructive\textquotedblright{}
interference phases $\text{mod}(\tilde{\phi}_{12},2\pi)=\pi$ when
the chirality of couplings is tuned between the BEC (UUEC) regime
and the BUEC regime, and moreover, the width of the reflection window
can be flexibly controlled {[}as shown in Fig. \ref{fig2_Txy}(a){]}.
The chiral setups thus have the merits of flexibility and tunability
in controlling photon transmission, offering the potential application
for sensing and optical switch \citep{4JiaWZ-AnAtom,24JiaWZ_NAtoms,AFK-chiral-23}.

For $N>2$, we consider the simplified model where the coupling strengths
at the $i$th coupling point $g_{Li}$ ($g_{Ri}$) for the left-propagating
(right-propagating) photons are uniquely different from those {[}assumed
to be identical to $g_{L}$ ($g_{R}$){]} of other ($N-1$) coupling
points. Note that in this case both the Lamb shift $\Delta_{ls}$
and the decay rates ($\Gamma_{L},\Gamma_{R}$) strongly depend on
the differences of the coupling strengths $\left(g_{L}-g_{Li}\right)$
and $\left(g_{R}-g_{Ri}\right)$, and moreover relate to the specific
position $z_{i}$ of the coupling point with distinct coupling strengths,
see Appendix B.

\begin{figure*}
\includegraphics[width=0.85\textwidth]{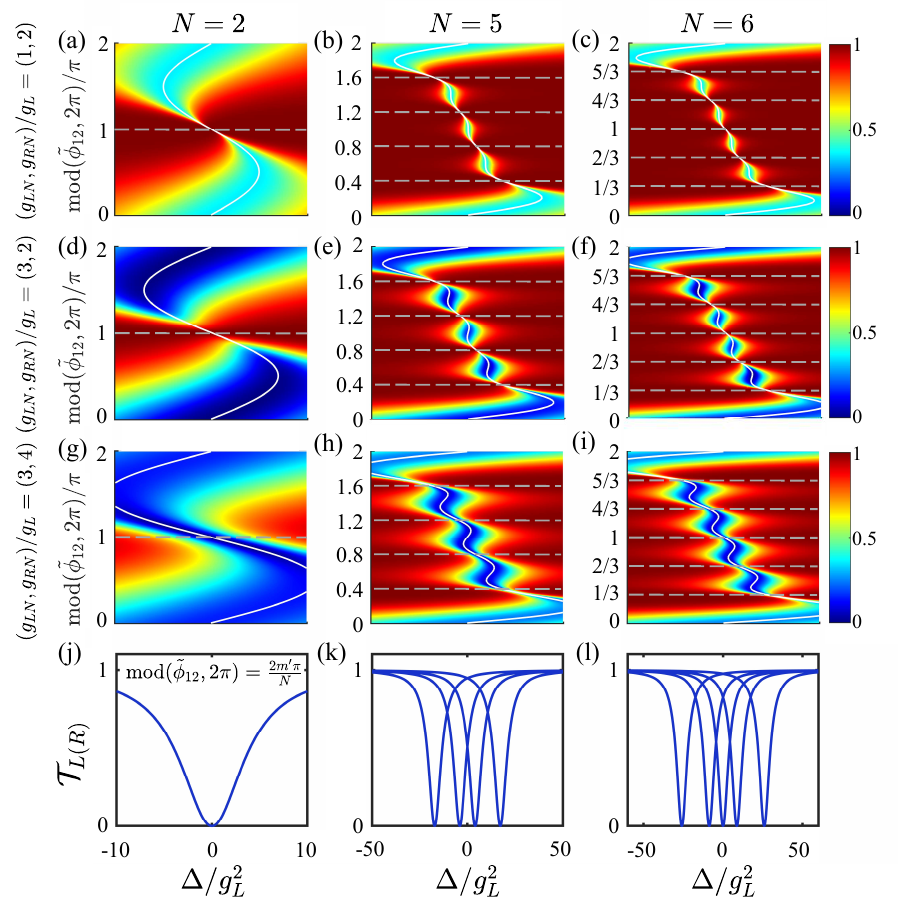} \caption{\label{Fig:T.vs.Delta_phi}Transmission probabilities $\mathcal{T}_{L(R)}$
versus $\Delta$ and $\text{mod}(\tilde{\phi}_{12},2\pi)$ for $N=2$,
5, 6, respectively. We consider the rightmost (i.e., $N$th) coupling
point which has the coupling strength $(g_{LN},g_{RN})/g_{L}$ different
to those of the others with $(g_{Lj},g_{Rj})/g_{L}=(1,2)$ ($j\protect\neq N$).
As indicated in Figs. \ref{fig2_Txy}(b)-\ref{fig2_Txy}(d), we set
$(g_{LN},g_{RN})/g_{L}=\{(1,2),(3,2),(3,4)\}$, which (from top to
bottom) correspond to the BEC {[}(a)-(c){]}, the UUEC {[}(d)-(f){]},
and the BUEC {[}(g)-(i){]} regime, respectively. The white solid lines
are used to label the solutions for $\Delta=\Delta_{ls}(\tilde{\phi}_{12})$,
and the horizontal gray dashed lines label the phases $\text{mod}(\tilde{\phi}_{12},2\pi)=2m^{\text{\ensuremath{\prime}}}\pi/N$
($m^{\text{\ensuremath{\prime}}}=1,2,...,N-1$). Panels (j)-(l) depict
transmission spectrum in the BUEC regime by fixing $\text{mod}(\tilde{\phi}_{12},2\pi)=2m^{\text{\ensuremath{\prime}}}\pi/N$,
where for $N$ being an even number, one can observe $\mathcal{T}_{L(R)}=0$
at the resonance $\Delta=0$ with $N-2$ dips symmetrically around
it. This can potentially be used for multi-channel photon routing.}
\end{figure*}

(1) For $\text{mod}(\tilde{\phi}_{12},2\pi)=0$, the Lamb shifts vanish
again, but the decay rates become dependent on the number of the coupling
points, see Appendix B. Then, the incident photon is totally reflected
at $\Delta=0$ if the chirality of the coupling strengths satisfies
\begin{equation}
\left(N-1\right)g_{L}+g_{Li}=\left(N-1\right)g_{R}+g_{Ri},\label{eq:Chiralcondition1}
\end{equation}
which reduces to $g_{L}+g_{Li}=g_{R}+g_{Ri}$ for $N=2$. In Fig.
\ref{fig2_Txy}(b) {[}Fig. \ref{fig2_Txy}(c){]}, we show $\mathcal{T}_{L(R)}[\Delta=0,\text{mod}(\tilde{\phi}_{12},2\pi)=0]$
as functions of $g_{Li}$ and $g_{Ri}$ (in units of $g_{L}$) with
$g_{R}/g_{L}=2$ and $N=2$ ($N=5$), and indicate $\mathcal{T}_{L(R)}=0$
(or $\mathcal{R}_{L(R)}=1$) corresponding to the condition {[}Eq.
(\ref{eq:Chiralcondition1}){]} by white dashed lines. Around the
resonance, $\mathcal{T}_{L(R)}(\Delta)$ under the condition of Eq.
(\ref{eq:Chiralcondition1}) possess the anti-Lorentzian line shape
with the FWHM $2\left[\left(N-1\right)g_{L}+g_{Li}\right]^{2}$ scaling
as $N^{2}$.

(2) For $\text{mod}(\tilde{\phi}_{12},2\pi)=2m^{\text{\ensuremath{\prime}}}\pi/N$,
it is interesting to see that $\Gamma_{L}\pm\Gamma_{R}$ are independent
of $N$ and are simply determined by the differences of the coupling
strengths $\left(g_{L}-g_{Li}\right)$ and $\left(g_{R}-g_{Ri}\right)$,
see Appendix B. As a result, the scattering behavior is similar to
the case of $N=2$, where an incident photon is perfectly transmitted
{[}$\mathcal{T}_{L(R)}(\Delta)\equiv1${]} if $g_{L}=g_{Li}$ or $g_{R}=g_{Ri}$,
regardless of the specific value of the detuning $\Delta$. In the
BUEC regime, when the coupling strengths of the $i$th coupling point
fulfill the condition 
\begin{equation}
|g_{L}-g_{Li}|=|g_{R}-g_{Ri}|,\label{eq:ChiralCondition2}
\end{equation}
the transmission probabilities reduce to 
\begin{equation}
\mathcal{T}_{L(R)}=\frac{\left(\Delta-\Delta_{ls}\right)^{2}}{\left(\Delta-\Delta_{ls}\right)^{2}+\left(g_{L}-g_{Li}\right)^{4}},
\end{equation}
which again possess the anti-Lorentzian line shape of the tunable
width $2(g_{L}-g_{Li})^{2}$ but with the center ($\mathcal{T}_{L(R)}=0$)
shifted to $\Delta=\Delta_{ls}$. In Fig. \ref{fig2_Txy}(d), we have
shown the density plot of $\mathcal{T}_{L(R)}(\Delta=\Delta_{ls})$
as functions of the chiral coupling $(g_{Lj},g_{Rj})/g_{L}$, and
indicated the UUEC regime by horizontal and vertical dotted lines
and the BUEC regime (with $|g_{L}-g_{Li}|=|g_{R}-g_{Ri}|$) by white
dashed line. The intersection of the white dashed line with the coordinate
$(g_{Lj},g_{Rj})/g_{L}=(1,2)$ corresponds to the BEC regime with
$\mathcal{T}_{L(R)}=1$. Furthermore, if we take $i=N$, i.e., the
coupling strengths of the leftmost (or rightmost) coupling point are
different from those of the others, the Lamb shift then reduces to
$\Delta_{ls}=\frac{N}{2}\left(g_{L}^{2}+g_{R}^{2}\right)\text{cot}(m^{\text{\ensuremath{\prime}}}\pi/N)$
(with $m^{\text{\ensuremath{\prime}}}=1,2,...,N-1$), which is exactly
the same to that of the BEC regime, see Eq. (\ref{eq:Lamb2_BEC}).
Note that $\Delta_{ls}$ is now \textit{independent} of the coupling
strengths \{$g_{LN}$, $g_{RN}$\} at the $N$th coupling point. Therefore,
by engineering the coupling strengths \{$g_{LN},g_{RN}$\}, the transmission
probabilities $\mathcal{T}_{L(R)}(\Delta)$ can be tuned between zero
and unity at the frequency detunings $\Delta=\Delta_{ls}(\tilde{\phi}_{12})$,
and the width of the ($N-1$) total reflection windows given by $2(g_{L}-g_{LN})^{2}$
can be flexibly controlled in the BUEC regime. This remarkable feature
is not present in the non-chiral setup with a large $N$ \citep{4JiaWZ-AnAtom},
as well as in a chiral setup (with only two coupling points) working
in the BEC (or UUEC) regime \citep{9DuL-,LiaoJQ,2DuL_lambda}.

In Fig. \ref{Fig:T.vs.Delta_phi}, we plot the transmission probabilities
$\mathcal{T}_{L(R)}$ as functions of the detuning $\Delta$ and the
two-point phase delay $\tilde{\phi}_{12}$ modulo $2\pi$ {[}i.e.,
$\text{mod}(\tilde{\phi}_{12},2\pi)${]} with $g_{R}/g_{L}=2$ and
$N=\{2,5,6$\}. We consider the set of coupling strengths $(g_{LN},g_{RN})/g_{L}=\{(1,2),(3,2),(3,4)\}$
for the $N$th coupling point, corresponding to the BEC, the UUEC,
and the BUEC regime, respectively. In all subfigures, the white solid
lines indicate $\Delta=\Delta_{ls}(\tilde{\phi}_{12})$. We first
look at the case of $N=2$ coupling points, for $\text{mod}(\tilde{\phi}_{12},2\pi)=0$,
the incident photon is partially reflected (i.e., $0<\mathcal{T}_{L(R)}<1$)
in the BEC regime with $\text{(}g_{LN},g_{RN})/g_{L}=(1,2)$ {[}Fig.
\ref{Fig:T.vs.Delta_phi}(a){]}, and nevertheless is totally reflected
(i.e $\mathcal{T}_{L(R)}=0$) in the UUEC regime with $\text{(}g_{LN},g_{RN})/g_{L}=(3,2)$
satisfying $g_{L}+g_{Li}=g_{R}+g_{Ri}$ {[}Fig. \ref{Fig:T.vs.Delta_phi}(d){]}.
In both cases, the incident photon is completely transmitted for $\text{mod}(\tilde{\phi}_{12},2\pi)=\pi$,
which is indicated by the gray dashed line. As a result, the two red
regions (corresponding to high transmission probabilities) connect
to each other around $\Delta=0$. In contrast, in the BUEC regime
with $\text{(}g_{LN},g_{RN})/g_{L}=(3,4)$ (satisfying $|g_{L}-g_{LN}|=|g_{R}-g_{RN}|$),
the transmission probabilities have the minimum $\mathcal{T}_{L(R)}=0$
precisely at $\Delta=0$ with $\text{mod}(\tilde{\phi}_{12},2\pi)=\pi$,
the two red regions are then disconnected {[}see Fig. \ref{Fig:T.vs.Delta_phi}(g){]}.
For $N=5$ ($N=6)$, as shown by the subfigures (b), (c), (e), and
(f), one then finds that the incident photon is completely transmitted
along the dashed lines {[}corresponding to $\text{mod}(\tilde{\phi}_{12},2\pi)=2m^{\text{\ensuremath{\prime}}}\pi/N$
$(m^{\text{\ensuremath{\prime}}}=1,2,...,N-1)${]} in the BEC and
the UUEC regime, and while is totally reflected at the detunings $\Delta\sim\Delta_{ls}=\frac{N}{2}\left(g_{L}^{2}+g_{R}^{2}\right)\text{cot}(\frac{m^{\text{\ensuremath{\prime}}}\pi}{N})$
in the BUEC regime {[}see Figs. \ref{Fig:T.vs.Delta_phi}(h) and \ref{Fig:T.vs.Delta_phi}(i){]}.
Moreover, in Figs. \ref{Fig:T.vs.Delta_phi}(j), \ref{Fig:T.vs.Delta_phi}(k)
and \ref{Fig:T.vs.Delta_phi}(l), we show the transmission spectrum
corresponding to the cut of the plots {[}Figs. \ref{Fig:T.vs.Delta_phi}(g),
\ref{Fig:T.vs.Delta_phi}(h) and \ref{Fig:T.vs.Delta_phi}(i){]} at
$\text{mod}(\tilde{\phi}_{12},2\pi)=2m^{\text{\ensuremath{\prime}}}\pi/N$
(indicated by the dashed lines), which possess the anti-Lorentzian
lineshape centered at $\Delta=\Delta_{ls}$ with a tunable width.
It is worth noting that $\Delta_{ls}$ is independent of the specific
values of the coupling strengths $(g_{Li},g_{Ri})$ for $i=N$, thus,
by setting $\Delta=\Delta_{ls}(\tilde{\phi}_{12}=\frac{2m^{\prime}\pi}{N})$,
the setup can potentially act as a multi-channel photon router, with
the frequencies of incident photons $\omega_{e}+\Delta_{ls}(\tilde{\phi}_{12})$
and the number of channels $N-1$ under the condition of $\text{mod}(\tilde{\phi}_{12},2\pi)\in(0,2\pi)$.

\begin{figure}
\includegraphics[width=1\columnwidth]{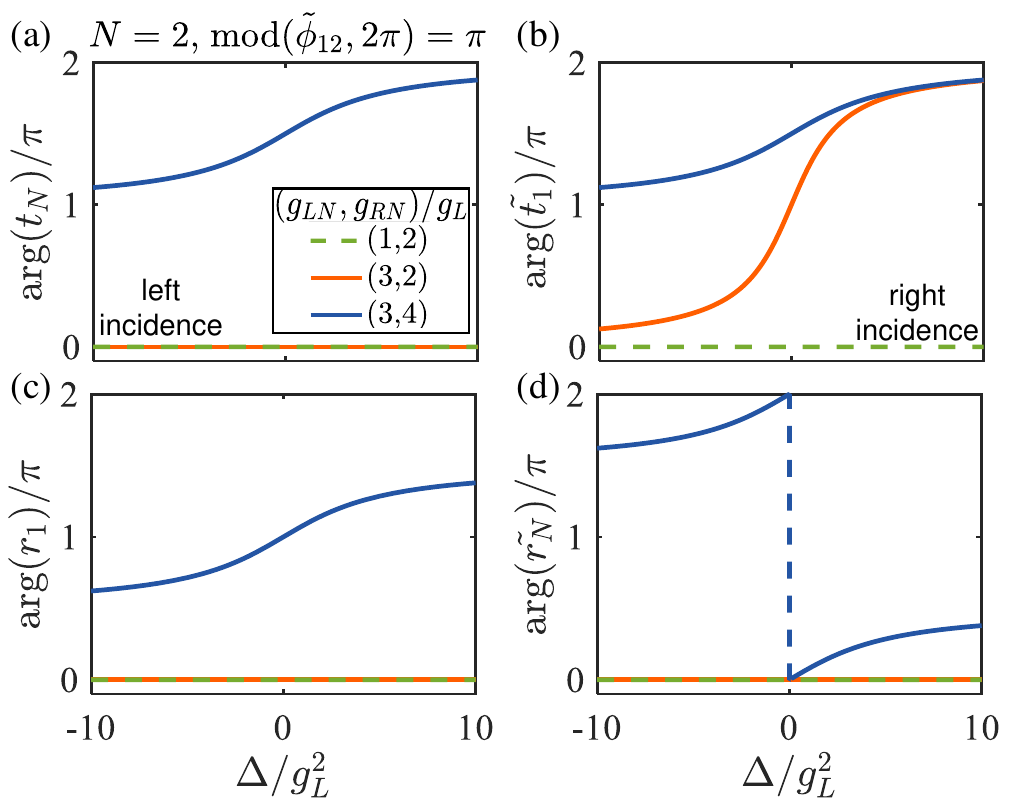}

\caption{\label{fig:Phases_N2}Phases of the transmission ($t_{N}$,$\tilde{t}_{1}$)
and reflection ($r_{1}$,$\tilde{r}_{N}$) coefficients versus $\Delta$
with $N=2$. The set of coupling strengths for the second coupling
point are $(g_{L2},g_{R2})/g_{L}=\{(1,2),(3,2),(3,4)\}$, and $(g_{L1},g_{R1})/g_{L}=(1,2)$.
The two-point propagating phase is set to $\text{mod}(\tilde{\phi}_{12},2\pi)=\pi$.}
\end{figure}

\begin{figure}
\includegraphics[width=1\columnwidth]{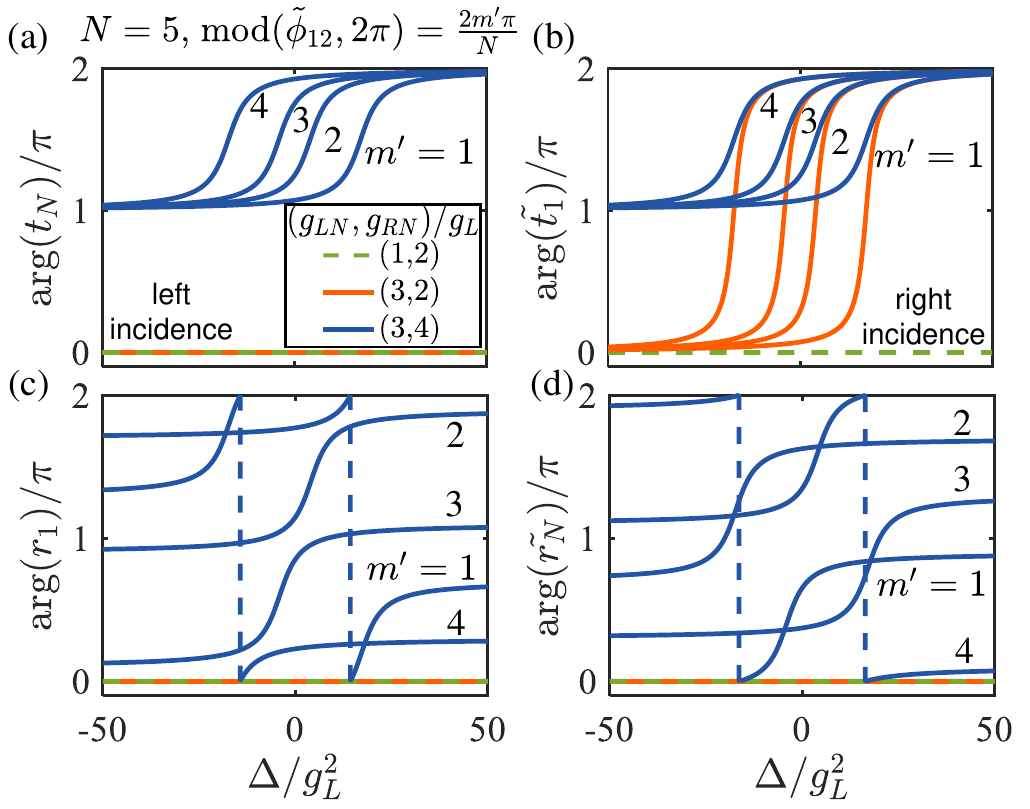}

\caption{\label{fig:Phases_N5}Phases of the transmission ($t_{N}$,$\tilde{t}_{1}$)
and reflection ($r_{1}$,$\tilde{r}_{N}$) coefficients versus $\Delta$
with $N=5$. We consider the rightmost (i.e., $N$th) coupling point
which has the coupling strengths $(g_{LN},g_{RN})/g_{L}=\{(1,2),(3,2),(3,4)\}$
differing from those $(g_{LN},g_{RN})/g_{L}=(1,2)$ ($j\protect\neq N$){]}
of the other coupling points. Here, the two-point propagating phases
are $\text{mod}(\tilde{\phi}_{12},2\pi)=2m^{\text{\ensuremath{\prime}}}\pi/N$
$(m^{\text{\ensuremath{\prime}}}=1,2,...,N-1)$.}
\end{figure}

Moreover, as shown in Fig. \ref{fig:Phases_N2} and Fig. \ref{fig:Phases_N5},
we examine the phases of the transmission coefficients {[}arg($t_{N}$)
and arg($\tilde{t}_{1}$){]} and the reflection coefficients {[}arg($r_{1}$)
and arg($\tilde{r}_{N}$){]} for the BEC, UUEC, and BUEC regimes,
respectively. The two-point propagating phases are set to $\text{mod}(\tilde{\phi}_{12},2\pi)=2m^{\text{\ensuremath{\prime}}}\pi/N$
$(m^{\text{\ensuremath{\prime}}}=1,2,...,N-1)$ (corresponding to
the horizontal dash lines in Fig. \ref{Fig:T.vs.Delta_phi}), where
the incident photon can be completely transmitted or totally reflected.

For $N=2$ and $(g_{LN},g_{RN})/g_{L}=\{(1,2),(3,2)\}$, we find that
$t_{N}=1$ and $r_{1}=0$ at any $\Delta$ for a left-incident photon,
leading to $\text{arg}(t_{N})=\text{arg}(r_{1})=0$, see Fig. \ref{fig:Phases_N2}(a)
and \ref{fig:Phases_N2}(c). For $(g_{LN},g_{RN})/g_{L}=(3,4)$, we
alternatively have $t_{N}=\frac{\Delta}{\Delta+i2\Gamma{}_{L}}$ with
arg$(t_{N})=-\text{atan}(2\Gamma_{L}/\Delta)$ and $r_{1}=\frac{-i\Gamma_{LR}}{\Delta+i2\Gamma{}_{L}}$
with arg$(r_{1})=\text{atan}(\Delta/2\Gamma_{L})$. In particular
for $\Delta\rightarrow0$, the photon is totally reflected and the
reflection coefficient $r_{1}$ picks up a phase of $\pi$. For a
right-incident photon, arg$(\tilde{t}_{1})$ for $(g_{LN},g_{RN})/g_{L}=(1,2)$
and $(g_{LN},g_{RN})/g_{L}=(3,4)$ are the same as those {[}i.e. $\text{arg}(t_{N})${]}
in the left-incident case {[}see Figs. \ref{fig:Phases_N2}(a) and
\ref{fig:Phases_N2}(b){]}; but for $(g_{LN},g_{RN})/g_{L}=(3,2)$,
one has $\tilde{t}_{1}=\frac{\Delta-i\Gamma_{L}}{\Delta+i\Gamma{}_{L}}$
and arg$(\tilde{t}_{1})=-2\text{atan}(\Gamma_{L}/\Delta)$, which
now depends on the frequency detuning $\Delta$. Moreover, for $(g_{LN},g_{RN})/g_{L}=(3,4)$,
the phase of the reflection coefficient becomes $\text{arg}(\tilde{r}_{N})=\text{arg}(r_{1})+\pi$
{[}see Fig. \ref{fig:Phases_N2}(d){]}, which implies that the right-incident
photon is reflected at $\Delta=0$ with $\text{arg}(\tilde{r}_{N})=2\pi$.
Thus, in the UUEC regime, one finds $\mathcal{T}_{L(R)}=1$ at any
$\Delta$, but a different transmission phase for the forward- and
backward-propagating photons. In the BUEC regime, one obtains $\mathcal{R}_{L(R)}=1$
at $\Delta=0$, but a different reflection phase for the forward-
and backward-propagating photons.

For $N=5$ and $\text{ mod}(\tilde{\phi}_{12},2\pi)=2m^{\text{\ensuremath{\prime}}}\pi/N$
$(m^{\text{\ensuremath{\prime}}}=1,2,...,N-1)$, the phases of the
transmission and reflection coefficients show similar features for
the forward- and the backward-propagating photons as those for $N=2$,
but now become dependent on $\tilde{\phi}_{12}$. For $(g_{LN},g_{RN})/g_{L}=(3,2)$,
the transmission phases are $\text{arg}(t_{N})=0$ and $\text{arg}(\tilde{t}_{1})=-2\text{atan}[\Gamma_{L}/(\Delta-\Delta_{ls})]$;
while for $(g_{LN},g_{RN})/g_{L}=(3,4)$, the transmission phases
are $\text{arg}(t_{N})=\text{arg}(\tilde{t}_{1})=-\text{atan}[2\Gamma_{L}/(\Delta-\Delta_{ls})]$.
As shown in Figs. \ref{fig:Phases_N5}(a) and \ref{fig:Phases_N5}(b),
when $m^{\text{\ensuremath{\prime}}}$ is varied, the curves corresponding
to the transmission phases are transversely displaced by $\Delta_{ls}(\frac{2m^{\text{\ensuremath{\prime}}}\pi}{N})$
in comparison with that of $N=2$ {[}where $\text{arg}(t_{N})=-\text{atan}(2\Gamma_{L}/\Delta)${]}.
On the other hand, the curves corresponding to the reflection phases
$\text{arg}(\tilde{r}_{N})$ and $\text{arg}(r_{1})$ are displaced
both transversely and longitudinally, see Figs. \ref{fig:Phases_N5}(c)
and \ref{fig:Phases_N5}(d). Consequently, the giant atom imprints
direction-dependent phases on the propagating photon uniquely for
the chiral setup in the UUEC and BUEC regimes.

\section{Nonreciprocal photon scattering with atomic spontaneous emission}

\begin{figure}
\includegraphics[width=1\columnwidth]{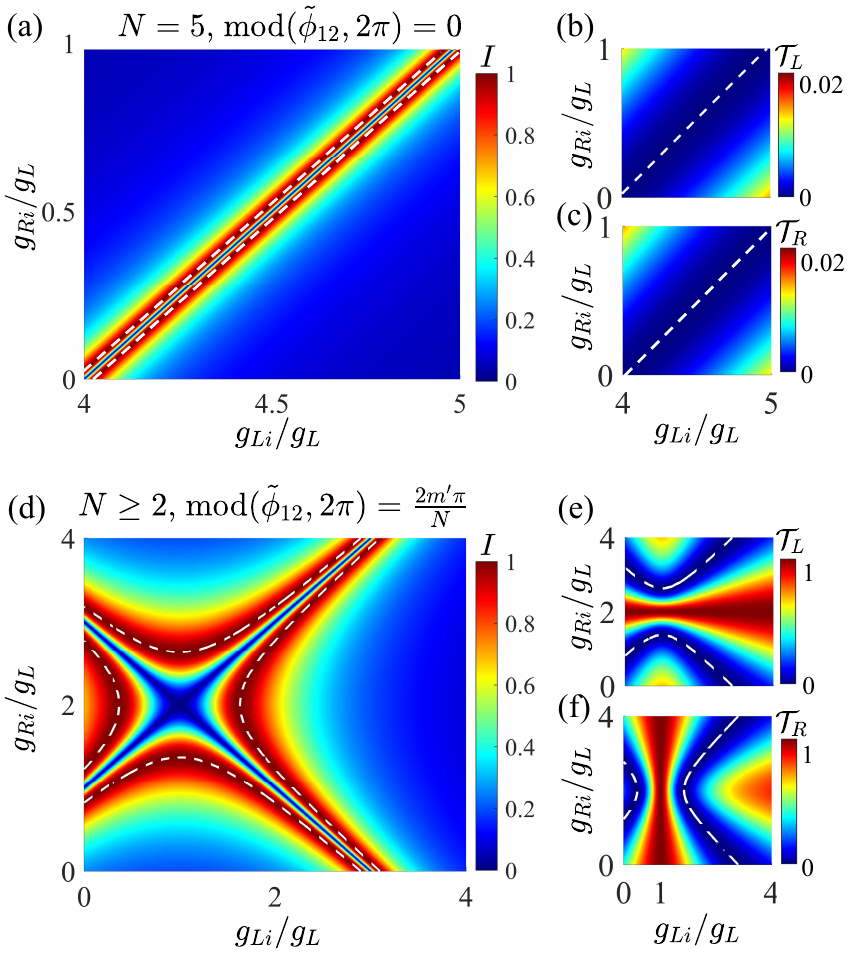}

\caption{\label{Fig:nonreciprocal_I} Nonreciprocal transport of a single photon
by including the spontaneous emission decay $\gamma/g_{L}^{2}=0.2$.
The contrast ratio $I$ versus $g_{Li}/g_{L}$ and $g_{Ri}/g_{L}$
for (a) $N=5$, $\text{mod}(\tilde{\phi}_{12},2\pi)=0$ and (d) $N\protect\geq2$,
$\text{mod}(\tilde{\phi}_{12},2\pi)=2m^{\text{\ensuremath{\prime}}}\pi/N$,
with the corresponding transmission probability $\mathcal{T}_{L}(\mathcal{T}_{R})$
for a left-incident (right-incident) photon being shown in (b) {[}(c){]}
and (e) {[}(f){]}. The white dashed lines in panels (a), (d) indicate
the optimal nonreciprocity condition with $I=1$. The white hyperbola
dashed lines in (b) {[}(c){]} and (e) {[}(f){]} are used to label
the transmission probabilities with $\mathcal{T}_{L(R)}=0$, corresponding
to $\gamma+\Gamma_{L}=\Gamma_{R}$ or $\gamma+\Gamma_{R}=\Gamma_{L}$.
Other parameters are the same as in Figs. \ref{fig2_Txy} (c)-(d).}
\end{figure}

We now take the atomic spontaneous decay into account, i.e.,\textbf{
$\gamma\neq0$}. Previously, we mentioned that nonreciprocal photon
scattering (i.e., $\mathcal{T}_{L}\ne\mathcal{T}_{R}$) can not occur
for the giant atom with uniformly symmetric coupling. Moreover, it
is remarkable that only reciprocal photon transport can be observed
with $\gamma=0$ despite the even or uneven chiral coupling. However,
we will see that, by engineering the chirality of the coupling strengths,
nonreciprocal photon transfer can be realized, due to the joint effect
of the chirality (i.e., $\Gamma_{L}\neq\Gamma_{R}$) and the atomic
spontaneous decay $\gamma\neq0$, and while the reflections are reciprocal
$\mathcal{R}_{L}=\mathcal{R}_{R}$, see Eqs. (\ref{eq:TR_left}) and
(\ref{eq:TR_right}). In other words, the atomic spontaneous decay
of the giant atom due to its coupling to the thermal environment and
the external decay due to the interfaced waveguide can cooperatively
lead to nonreciprocal photon transport.

As an example, for $\gamma=\Gamma_{L}-\Gamma_{R}$ and $\Delta=\Delta_{ls}$,
a photon incident from the left will be partially transmitted with
the probability 
\begin{equation}
\mathcal{T}_{L}=\left(1-\frac{\Gamma_{R}}{\Gamma_{L}}\right)^{2},
\end{equation}
while a photon incident from the right will be completely isolated,
i.e., $\text{ \ensuremath{\mathcal{T}_{R}}}=0$. For a generic situation,
the nonreciprocity in transmission can be described by a contrast
ratio defined as 
\begin{equation}
I=\left|\frac{\mathcal{T}_{L}-\mathcal{T}_{R}}{\mathcal{T}_{L}+\mathcal{T}_{R}}\right|=\frac{2\gamma\left|\Gamma_{L}-\Gamma_{R}\right|}{\left(\Delta-\Delta_{ls}\right)^{2}+\gamma^{2}+\left(\Gamma_{L}-\Gamma_{R}\right)^{2}}.\label{eq:I}
\end{equation}
Thus, the nonreciprocity can only be observed when both the conditions
$\gamma\neq0$ and $\Gamma_{L}\neq\Gamma_{R}$ are simultaneously
satisfied. By setting $\Delta=\Delta_{ls}$, one achieves the optimal
nonreciprocity $I=1$ with $\gamma=|\Gamma_{L}-\Gamma_{R}|,$ where
the spontaneous decay rate $\gamma$ cancels out the difference between
the effective decay rates in the forward and backward direction. For
$\Delta\neq\Delta_{ls}$, the contrast ratio is always less than one.
Note that here the nonreciprocity is induced by the chirality of couplings,
in contrast to that induced by non-Markovian retardation effects \citep{LiaoJQ}
and by synthetic gauge fields \citep{9DuL-}.

In Figs. \ref{Fig:nonreciprocal_I}(a) and \ref{Fig:nonreciprocal_I}(d),
we plot the contrast ratio $I$ as functions of the coupling strengths
$(g_{Li},g_{Ri})/g_{L}$ for $g_{R}/g_{L}=2$, $\gamma/g_{L}^{2}=0.2$,
(a)-(c) $N=5$, and (d)-(f) $N\geq2$. Correspondingly, we show the
forward (backward) transmission probabilities $\mathcal{T}_{L}(\Delta=\Delta_{ls})$
{[}$\mathcal{T}_{R}(\Delta=\Delta_{ls})${]} for a left-incident (right-incident)
photon in Fig. \ref{Fig:nonreciprocal_I}(b) {[}Fig. \ref{Fig:nonreciprocal_I}(c){]}
with $\text{mod}(\tilde{\phi}_{12},2\pi)=0$ and Fig. \ref{Fig:nonreciprocal_I}(e)
{[}Fig. \ref{Fig:nonreciprocal_I}(f){]} with $\text{mod}(\tilde{\phi}_{12},2\pi)=2m^{\text{\ensuremath{\prime}}}\pi/N$.
As discussed above, the optimal nonreciprocity ($I=1$) condition
for $\text{mod}(\tilde{\phi}_{12},2\pi)=0$ is 
\begin{equation}
\gamma=\pm\frac{\left[\left(N-1\right)g_{L}+g_{Li}\right]^{2}-\left[\left(N-1\right)g_{R}+g_{Ri}\right]^{2}}{2},
\end{equation}
and alternatively for $\text{mod}(\tilde{\phi}_{12},2\pi)=2m^{\text{\ensuremath{\prime}}}\pi/N$
($m^{\text{\ensuremath{\prime}}}=1,2,...,N-1$) is 
\begin{equation}
\gamma=\pm\frac{\left(g_{Li}-g_{L}\right)^{2}-\left(g_{Ri}-g_{R}\right)^{2}}{2},\label{eq:NPconditon2}
\end{equation}
corresponding to the hyperbola branches indicated by the white dashed
lines in Fig. \ref{Fig:nonreciprocal_I}(a) and Fig. \ref{Fig:nonreciprocal_I}(d).
Note that for $\text{mod}(\tilde{\phi}_{12},2\pi)=0$, one can only
observe the tail of hyperbola branches since we consider the coupling
strengths being positive values; when physical parameters (coupling
strengths) are extended to the complex plane, the full hyperbola branches
will be clearly seen in the lower left quadrant. However, the optimal
nonreciprocity is achieved at the expense of low transmission probabilities
$\mathcal{T}_{L(R)}$ {[}shown in Fig. \ref{Fig:nonreciprocal_I}(b)
and Fig. \ref{Fig:nonreciprocal_I}(c){]}. For $\text{mod}(\tilde{\phi}_{12},2\pi)=2m^{\text{\ensuremath{\prime}}}\pi/N$,
we previously show in the UUEC regime $\mathcal{T}_{L}=\mathcal{T}_{R}\equiv1$
along $g_{Ri}/g_{R}=1$ ($g_{Li}/g_{L}=1$) for $\gamma=0$, in contrast
to that, when $\gamma\neq0$, there emerges an avoided crossing at
the BEC point $(g_{Li},g_{Ri})/g_{L}=(1,2)$ with a gap $\sim2\sqrt{2\gamma}$
of the coupling strength $g_{Ri}/g_{R}$ ($g_{Li}/g_{L}$) for $\mathcal{T}_{L}$
($\mathcal{T}_{R}$) in Fig. \ref{Fig:nonreciprocal_I}(e) {[}Fig.
\ref{Fig:nonreciprocal_I}(f){]}. Here, the optimal nonreciprocity
{[}Eq. (\ref{eq:NPconditon2}){]} corresponds to the total reflection
of the incident photon $\mathcal{T}_{L}=0$ or $\mathcal{T}_{R}=0$,
as indicated by the white dashed lines in Fig. \ref{Fig:nonreciprocal_I}(e)
and Fig. \ref{Fig:nonreciprocal_I}(f). In particular, by considering
the UUEC regime (i.e., $\Gamma_{L}=0$ or $\Gamma_{R}=0$), we obtain
the pefect nonreciprocity $I=1$ with $\mathcal{T}_{L}=0$ ($\mathcal{T}_{R}=1)$
or $\mathcal{T}_{L}=1$ ($\mathcal{T}_{R}=0)$ for a left-incident
(right-incident) photon, since the chiral coupling $g_{Ri}$ ($g_{Li}$)
at the $i$th coupling point is counteracted by the spontaneous decay
according to $g_{Ri}=g_{R}\pm\sqrt{2\gamma}$ ($g_{Li}=g_{L}\pm\sqrt{2\gamma}$).
Extending to a generic case of chiral coupling, including the BEC,
the UUEC, and the BUEC regime, one finally observes $I=1$ on the
hyperbola branches in Fig. \ref{Fig:nonreciprocal_I}(d). This gives
a more complete picture about the chiral coupling induced nonreciprocity,
which is not limited to the special cases with two-point asymmetric
couplings \citep{9DuL-,LiaoJQ,2DuL_lambda}.

\section{The Non-Markovian regime}

So far, we consider only the Markovian regime, namely, the time for
light to travel between the leftmost and the rightmost coupling points
is much less than the characteristic relaxation time $\tau_{1N}\ll\tilde{\Gamma}^{-1}$.
For the giant atom with $N$ equally spaced coupling points, where
$\tilde{\Gamma}$ is on the same order of $N^{2}g_{L}^{2}$ (with
regard to the uniformly symmetric coupling) \citep{non-3GiantAcoustic,4JiaWZ-AnAtom,Ask2019},
the system in the non-Markovian regime implies $\tau_{1N}=(N-1)\tau_{12}$
being comparable to or even larger than $\tilde{\Gamma}^{-1}$. Moreover,
by considering the bandwidth of the order $|\Delta|\sim\tilde{\Gamma}$
or $|\Delta|/g_{L}^{2}\sim N^{2}$, the effect of $\Delta\tau_{12}$
on the propagating phase $\phi_{12}=\Delta\tau_{12}+\tilde{\phi}_{12}$
(and therefore the non-Markovian retardation effect) can not be ignored
\citep{2DuL_lambda,LiaoJQ}.

\begin{figure}
\includegraphics[width=1\columnwidth]{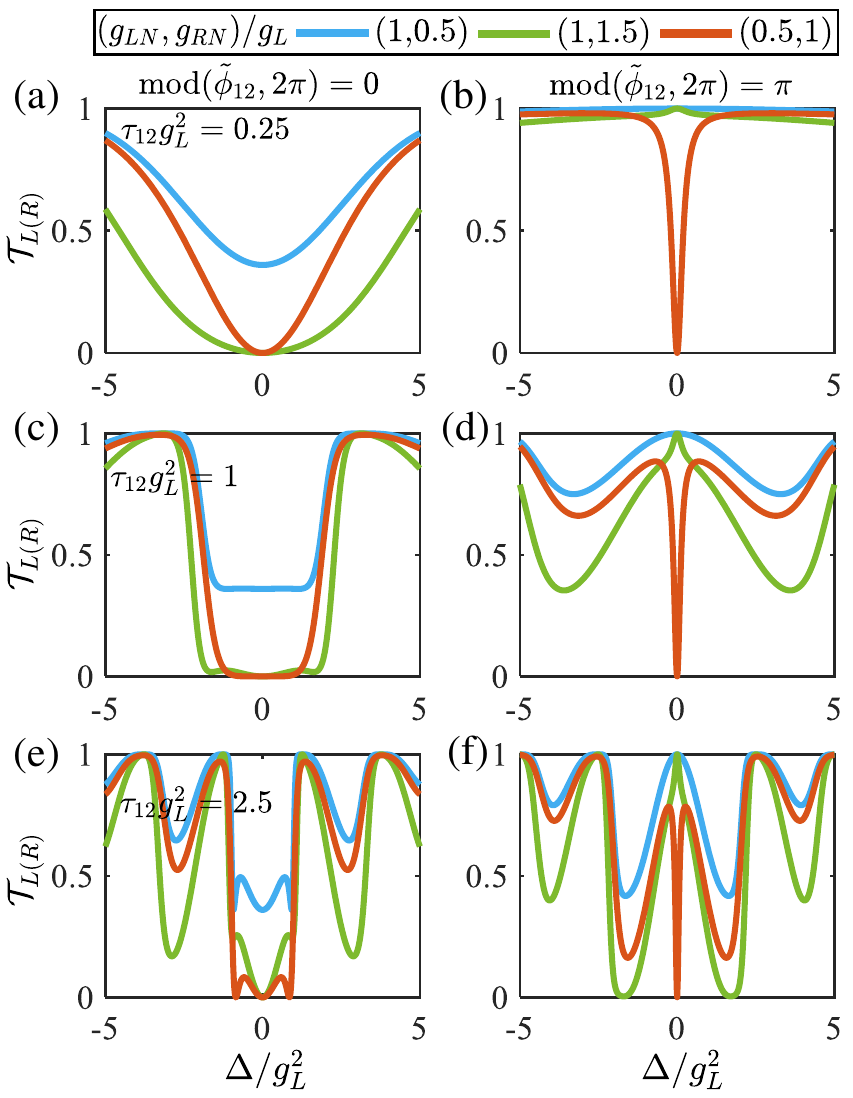} \caption{\label{Fig:T_NMarkov_N2}Transmission spectra in the non-Markovian
regime with $N=2$ and $\gamma=0$, where (a), (c), (e) $\text{mod}(\tilde{\phi}_{12},2\pi)=0$
(the left panels), and (b), (d), (f) $\text{mod}(\tilde{\phi}_{12},2\pi)=\pi$
(the right panels). The photon propagation time is set as $g_{L}^{2}\tau_{12}=\{0.25,1,2.5\}$,
corresponding to the weak, intermediate, and deep non-Markovian regimes,
respectively. Coupling strengths for the two coupling points are $(g_{L1},g_{R1},g_{L2},g_{R2})/g_{L}=\{(1,0.5,1,0.5),(1,0.5,1,1.5),(1,0.5,0.5,1)\}$,
which correspond to the BEC (blue), the UUEC (green), and the BUEC
(red) regime, respectively.}
\end{figure}

In Fig. \ref{Fig:T_NMarkov_N2}, we show transmission spectra for
$N=2$ coupling points as functions of $\Delta/g_{L}^{2}$ for $g_{L}^{2}\tau_{12}=0.25,1,2.5$,
respectively. We consider the set of coupling strengths with $(g_{L1},g_{R1},g_{L2},g_{R2})/g_{L}=(1,0.5,1,0.5)$,
$(g_{L1},g_{R1},g_{L2},g_{R2})/g_{L}=(1,0.5,1,1.5)$, and $(g_{L1},g_{R1},g_{L2},g_{R2})/g_{L}=(1,0.5,0.5,1)$,
according to the specific chiral conditions Eqs. (\ref{eq:Chiralcondition1})
and (\ref{eq:ChiralCondition2}). For $g_{L}^{2}\tau_{12}=0.25$,
the weak non-Markovian effect starts to influence the transmission
spectra. As shown in Fig. \ref{Fig:T_NMarkov_N2}(a), for $\text{mod}(\tilde{\phi}_{12},2\pi)=0$,
all of the scattering spectrum have a single minimum at $\Delta=0$
but with a different width when the coupling strengths $(g_{L2},g_{R2})/g_{L}$
are varied. For $\text{mod}(\tilde{\phi}_{12},2\pi)=\pi$, as $|\Delta|/g_{L}^{2}$
increases, the transmission probabilities in the UUEC {[}with $(g_{L1},g_{R1},g_{L2},g_{R2})/g_{L}=(1,0.5,1,1.5)${]}
and the BEC {[}with $(g_{L1},g_{R1},g_{L2},g_{R2})/g_{L}=(1,0.5,1,0.5)${]}
regime decline weakly from $\mathcal{T}_{L(R)}(\Delta=0)=1$, while
in the BUEC regime {[}with $(g_{L1},g_{R1},g_{L2},g_{R2})/g_{L}=(1,0.5,0.5,1)${]},
$\mathcal{T}_{L(R)}(\Delta)$ quickly transits from zero to almost
one for being out of resonance {[}see Fig. \ref{Fig:T_NMarkov_N2}(b){]}.
For $g_{L}^{2}\tau_{12}=1$, the system enters the intermediate non-Markovian
regime. For $\text{mod}(\tilde{\phi}_{12},2\pi)=0$, there emerges
a wide frequency interval where $\mathcal{T}_{L(R)}$ remain approximately
unchanged, see Fig. \ref{Fig:T_NMarkov_N2}(c). In particular, $\mathcal{T}_{L(R)}\simeq0$
in the frequency interval for $(g_{L2},g_{R2})/g_{L}=(0.5,1)$ (corresponding
to the BUEC regime), which is referred to as a photonic band gap with
the bandwidth scaling as $|\Delta|/g_{L}^{2}\propto N^{2}$ \citep{4JiaWZ-AnAtom}.
Note that this feature is previously found in a non-chiral setup \citep{4JiaWZ-AnAtom},
but with a large number of coupling points. For $\text{mod}(\tilde{\phi}_{12},2\pi)=\pi$,
as shown in Fig. \ref{Fig:T_NMarkov_N2}(d), $\mathcal{T}_{L(R)}(\Delta)$
start to oscillate and additional minima at around $\Delta/g_{L}^{2}\sim3$
can be found. In the deep non-Markovian regime with $g_{L}^{2}\tau_{12}=2.5$
{[}as shown in Figs. \ref{Fig:T_NMarkov_N2}(e) and \ref{Fig:T_NMarkov_N2}(f){]},
the flat lineshape in the band gap splits and exhibits side valleys
near $\Delta=0$ for $\text{mod}(\tilde{\phi}_{12},2\pi)=0$, and
only in the case of $(g_{L2},g_{R2})/g_{L}=(0.5,1)$ with respect
to the BUEC regime, we observe $\mathcal{T}_{L(R)}=0$ at the two
side valleys; besides, for $(g_{L2},g_{R2})/g_{L}=(1,1.5)$ and $(g_{L2},g_{R2})/g_{L}=(1,0.5)$,
$\mathcal{T}_{L(R)}(\Delta)$ oscillate between 0 and 1 as $|\Delta|$
increases whether $\text{mod}(\tilde{\phi}_{12},2\pi)=0$ or $\text{mod}(\tilde{\phi}_{12},2\pi)=\pi$.

\begin{figure}
\includegraphics[width=1\columnwidth]{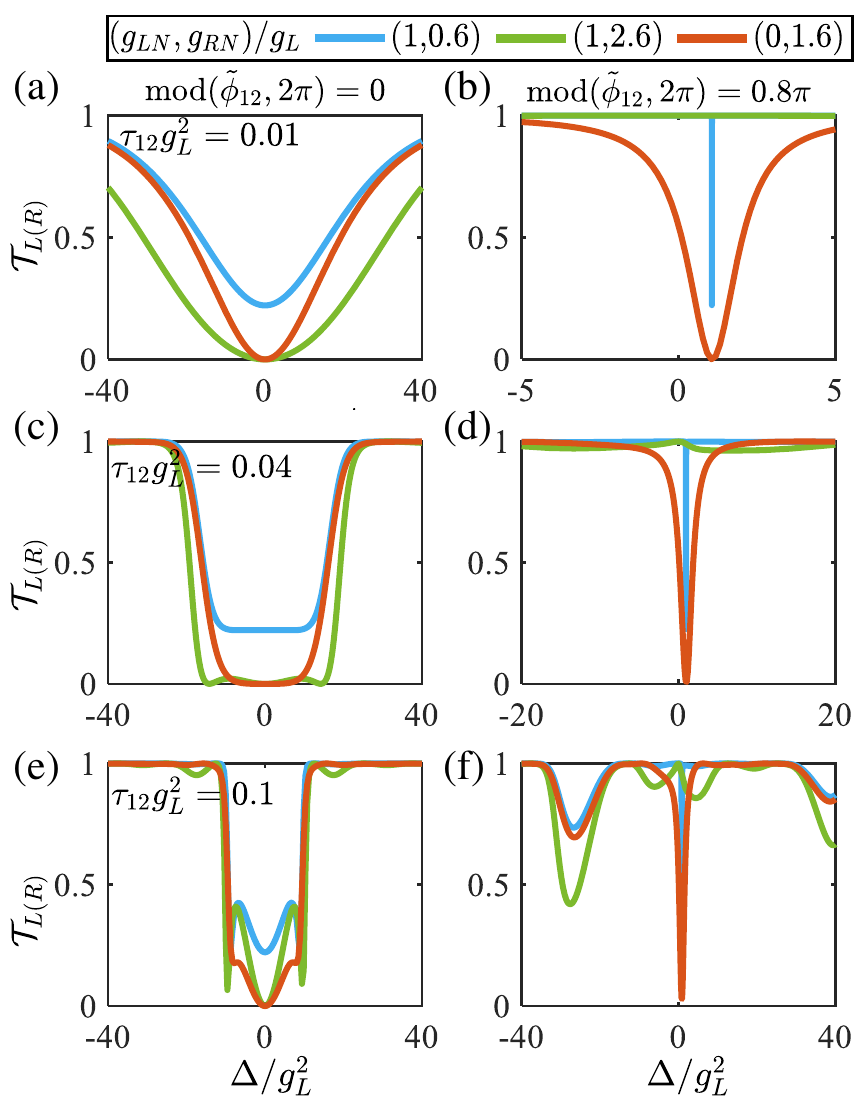}

\caption{\label{Fig:T_NMarkov_N5}Transmission spectrum in the non-Markovian
regime with $N=5$ and $\gamma=0$, where (a), (c), (e) $\text{mod}(\tilde{\phi}_{12},2\pi)=0$,
and (b), (d), (f) $\text{mod}(\tilde{\phi}_{12},2\pi)=0.8\pi$ {[}i.e.,
$\text{mod}(\tilde{\phi}_{12},2\pi)=2m^{\text{\ensuremath{\prime}}}\pi/N$
with $m^{\text{\ensuremath{\prime}}}=2$ without loss of generality{]}.
We set the coupling strengths of the $N$th coupling point as $(g_{LN},g_{RN})/g_{L}=\{(1,0.6),(1,2.6),(0,1.6)\}$
and that of the others as $(g_{Lj},g_{Rj})/g_{L}=(1,0.6)$ ($j\protect\neq N$),
which correspond to the BEC (blue), the UUEC (green), and the BUEC
(red) regime, respectively. Moreover, the photon propagation time
is set as $g_{L}^{2}\tau_{12}=\{0.01,0.04,0.1\}$, corresponding to
the weak, intermediate, and deep non-Markovian regimes, respectively.}
\end{figure}

In Fig. \ref{Fig:T_NMarkov_N5}, we show the transmission spectra
as functions of $\Delta/g_{L}^{2}$ with $N=5$ for $g_{L}^{2}\tau_{12}=0.01,0.04,\text{ and }0.1$,
which correspond to the weak, intermediate and deep non-Markovian
regimes, respectively. We consider the propagating time $\tau_{1N}\sim\left(N^{2}g_{L}^{2}\right)^{-1}$
from the first to the last coupling point, and the frequency bandwidth
is set to $|\Delta|\sim N^{2}g_{L}^{2}$. Furthermore, we assume that
the set of coupling strengths for the $N$th coupling point are $(g_{LN},g_{RN})/g_{L}=\{(1,0.6),(1,2.6),(0,1.6)\}$,
and the coupling strengths for the other coupling points are $(g_{Lj},g_{Rj})/g_{L}=(1,0.6)$,
corresponding to the three chiral coupling regimes discussed above.
Note that the coupling strengths $(g_{LN},g_{RN})/g_{L}=(0,0.6)$
satisfy both the conditions $g_{LN}-g_{RN}=\left(1-N\right)\left(g_{L}-g_{R}\right)$
and $|g_{LN}-g_{L}|=|g_{R}-g_{RN}$|, while the coupling strengths
$(g_{LN},g_{RN})/g_{L}=(1,2.6)$ satisfy $g_{LN}-g_{RN}=\left(1-N\right)\left(g_{L}-g_{R}\right)$
only. For $\text{mod}(\tilde{\phi}_{12},2\pi)=0$, as shown by Figs.
\ref{Fig:T_NMarkov_N5}(a), \ref{Fig:T_NMarkov_N5}(c) and \ref{Fig:T_NMarkov_N5}(e),
the transmission spectra with $g_{L}^{2}\tau_{12}=\{0.01,0.04,0.1\}$
show oscillation behaviors similar to those of $N=2$ in the same
non-Markovian regime, and as $|\Delta|$ increases, $\mathcal{T}_{L(R)}(\Delta)$
saturate to unity after a few oscillations in the intermediate and
deep non-Markovian regimes. For $\text{mod}(\tilde{\phi}_{12},2\pi)=2m^{\text{\ensuremath{\prime}}}\pi/N$,
the lineshape of the scattering spectra becomes asymmetric for an
odd number of coupling points, see Figs. \ref{Fig:T_NMarkov_N5}(b),
\ref{Fig:T_NMarkov_N5}(d) and \ref{Fig:T_NMarkov_N5}(f), where we
have taken $m^{\text{\ensuremath{\prime}}}=2$ {[}i.e., $\text{mod}(\tilde{\phi}_{12},2\pi)=0.8\pi${]}
as an example. In particular, since the Lamb shift $\Delta_{ls}$
is nonvanishing for $\text{mod}(\tilde{\phi}_{12},2\pi)=2m^{\text{\ensuremath{\prime}}}\pi/N$,
the time delay $\sim\Delta\tau_{12}$ will lead to a shift of the
dip {[}corresponding to $\mathcal{T}_{L(R)}(\Delta_{ls})\rightarrow0${]}
for the BUEC regime {[}i.e., $(g_{LN},g_{RN})/g_{L}=(0,1.6)${]},
while in comparison, $\mathcal{T}_{L(R)}$ remain unity at resonance
{[}i.e., $\mathcal{T}_{L(R)}(\Delta=0)\equiv1${]} for the UUEC {[}$(g_{LN},g_{RN})/g_{L}=(1,2.6)${]}
and BEC {[}$(g_{LN},g_{RN})/g_{L}=(1,0.6)${]} regimes. Note that,
since there may not exist an appropriate detuning $\Delta$ fulfilling
both $\Gamma_{L}=\Gamma_{R}$ and $\Delta=\Delta_{ls}$, $\mathcal{T}_{L(R)}$
in the BEC regime suffer from a sudden jump to $\mathcal{T}_{L(R)}(\Delta_{ls})=\left[(\Gamma_{L}-\Gamma_{R})/(\Gamma_{L}+\Gamma_{R})\right]^{2}$
at the detuning $\Delta=\Delta_{ls}$ {[}see the blue curve in Figs.
\ref{Fig:T_NMarkov_N5}(b), \ref{Fig:T_NMarkov_N5}(d), and \ref{Fig:T_NMarkov_N5}(f){]},
which does not occur in the case of $N=2$. In general, for $\text{mod}(\tilde{\phi}_{12},2\pi)=2m^{\text{\ensuremath{\prime}}}\pi/N$,
the oscillation amplitudes of $\mathcal{T}_{L(R)}(\Delta)$ become
larger when the system enters the intermediate and deep non-Markovian
regimes.

\section{Experimental feasibility and CONCLUSION}

The giant atom model under consideration can be found in Refs. \citep{AFK_experiment-Kannan2020}
and \citep{7Engineering}, where the three- and six-coupling-point
architecture are realized with non-chiral waveguides and are used
to demonstrate decoherence-free interaction \citep{AFK_experiment-Kannan2020}
or electromagnetically induced transparency \citep{7Engineering}.
In these experiments, a frequency-tunable transmon qubit capacitively
couples to a meandering microwave TL at multiple points, where the
number of coupling points is limited by the device geometry and the
physical size of the transmon qubit. The qubit can be regarded as
a two-level system when the atom-waveguide coupling strength $\sim g_{L}$
(of a few MHz) is much smaller than the level anharmonicity (with
the maximum value a few hundred MHz) \citep{AFK-chiral-69}, and the
atomic decay rate is on the order of $\sim0.1$ MHz \citep{AFK_experiment-Kannan2020}.
Moreover, the capacitive coupling at each connection point can be
tuned by mediating the qubit-field interaction with SQUIDs (i.e. superconducting
quantum interference devices) and tuning the magnetic fluxes threading
them \citep{7Engineering}. For the two-point propagating phase $\phi_{12}=\omega_{e}|z_{2}-z_{1}|/v$
around the qubit frequency $\omega_{e}$, where $v$ is the speed
of light in the waveguide, a phase $\phi_{12}$ of $2\pi$ corresponds
to a two-point distance $|z_{2}-z_{1}|=2\pi v/\omega_{e}\sim$ 20
mm, which can be controlled with great precision \citep{7Engineering}.
On the other hand, chiral quantum optics with photonic reservoir has
been studied with many architectures \citep{Lodahl2017}, e.g. photonic-crystal
waveguides or optical nanofibers with transversely confined light
\citep{AFK_ArtificialAtom,Soellner2015}, and microwave TLs with circulators
\citep{AFK-chiral-46,AFK-chiral-47,AFK-chiral-48,AFK-chiral-49}.
Considering that the giant artificial atoms demonstrated in Refs.
\citep{AFK_experiment-Kannan2020} and \citep{7Engineering} are based
on coupling to the microwave TLs, the scattering phenomena unique
to giant atoms chirally coupled with waveguides may be demonstrated
by using microwave circulators to provide the chirality \citep{AFK-chiral,AFK-chiral-46,AFK-chiral-47,AFK-chiral-48,AFK-chiral-49}.
Although the non-Markovian effect has not yet been observed in the
microwave-photons-based systems so far, it can be alternatively demonstrated
with the SAW-based systems \citep{non-4non-exponentialDecay}, or
by increasing the atom-waveguide coupling with a well-designed superconducting
flux qubit \citep{FornDiaz2017}

In conclusion, we have studied the photon scattering spectra by a
giant atom, which is chirally coupled to a waveguide at multiple equally
spaced points. The chirality is categorized in terms of the evenness
of coupling strengths in the left and right propagation directions,
and is divided into three regimes, i.e., the BEC, UUEC, and BUEC regimes.
In the Markovian limit and for the two-point propagating phase being
$\text{mod}(\tilde{\phi}_{12},2\pi)=0$, the incident photon can be
totally reflected at resonance and the transmission spectra possess
the anti-Lorentzian line shape with the width depending on the number
of coupling points $N$. For the two-point propagating phase satisfying
mod($\tilde{\phi}_{12},2\pi)=2m^{\prime}\pi/N$, the incident photon
is fully transmitted at $\Delta=\Delta_{ls}$ in the BEC and the UUEC
regimes, but can be totally reflected in the BUEC regime, where, interestingly,
the transmission probabilities can be flexibly tuned at ($N-1$) fixed
frequencies by engineering the chirality of couplings, allowing for
multiple-channel photon routing; moreover, the giant-atom-waveguide
coupling can imprint direction-dependent phases on the scattering
coefficients. Due to the chiral coupling, the maximum nonreciprocity
in photon scattering (corresponding to the contrast ratio of unity)
can be achieved when the atomic spontaneous decay is taken into account,
differing from the non-Markovianity induced effect in other chiral
waveguide QED systems \citep{2DuL_lambda,LiaoJQ}. The non-Markovian
retardation effect manifested by the scattering spectra reflects itself
mainly in the photonic band gap and the oscillatory behavior between
total reflection and full transmission, especially in the intermediate
and deep non-Markovian regimes, respectively. The giant-atom-waveguide
system with chiral coupling thus offers flexible ways for single-photon
routing.
\begin{acknowledgments}
H.W. acknowledges support from the National Natural Science Foundation
of China (NSFC) under Grants No. 11774058 and No. 12174058. Y.L. acknowledges
support from the NSFC under Grants No. 12074030 and No. 12274107.
\end{acknowledgments}

\section*{Appendix A Solution of transmission and reflection coefficients}

Inserting Eqs. (\ref{eq:Hamiltonian}) and (\ref{eq:eigenstate})
into the stationary Schrödinger equation $H|\psi\rangle=E|\psi\rangle$,
we find that the probability amplitudes obey the following relations:
\begin{align}
Ec_{gL}\left(z\right)= & (\omega_{0}+iv_{g}\frac{\partial}{\partial z})c_{gL}\left(z\right)\nonumber \\
 & +\stackrel[j=1]{N}{\sum}\delta(z-z_{j})\sqrt{v_{g}}g_{Lj}e^{ik_{0}z}c_{e0},\nonumber \\
Ec_{gR}\left(z\right)= & (\omega_{0}-iv_{g}\frac{\partial}{\partial z})c_{gR}\left(z\right)\nonumber \\
 & +\stackrel[j=1]{N}{\sum}\delta(z-z_{j})\sqrt{v_{g}}g_{Rj}e^{-ik_{0}z}c_{e0},\nonumber \\
(\Delta+i\gamma)c_{e0}= & \int dz\stackrel[j=1]{N}{\sum}\delta(z-z_{j})\sqrt{v_{g}}[g_{Lj}e^{-ik_{0}z}c_{gL}\left(z\right)\nonumber \\
 & +g_{Rj}e^{ik_{0}z}c_{gR}\left(z\right)],\label{eq:Schrodinger}
\end{align}
where $\Delta=\omega(k)-\omega_{e}$ is the detuning between the incident
photons and the atomic transition $|g\rangle\leftrightarrow|e\rangle$.
Now suppose a photon is incident from the left port and the atom is
initially in $|g\rangle$, the probability amplitudes, due to the
$\delta$-function potential effect of the atom at the coupling point,
can be formed as 
\begin{align}
c_{gR}\left(z\right)= & e^{i(k-k_{0})z}[\Theta\left(z_{1}-z\right)\nonumber \\
 & +\stackrel[j=1]{N-1}{\sum}t_{j}\Theta\left(z-z_{j}\right)\Theta\left(z_{j+1}-z\right)\nonumber \\
 & +t_{N}\Theta\left(z-z_{N}\right)],\nonumber \\
c_{gL}\left(z\right)= & e^{-i(k-k_{0})z}[r_{1}\Theta(z_{1}-z)\nonumber \\
 & +\stackrel[j=2]{N}{\sum}r_{j}\Theta(z-z_{j-1})\Theta(z_{j}-z)],\label{eq:LeftAmplitude}
\end{align}
where $t_{j}$ ($r_{j}$) is the transmission (reflection) coefficient
for the $j$th coupling point, $t_{N}$ ($r_{1}$) is the transmission
(reflection) coefficient for the last (first) coupling point, and
$\Theta(z-z_{j})$ are the Heaviside step function. While for the
right-incident case, the probability amplitudes alternatively take
the form 
\begin{align}
c_{gR}\left(z\right)= & e^{i(k-k_{0})z}[\stackrel[j=1]{N-1}{\sum}\tilde{r}_{j}\Theta(z-z_{j})\Theta(z_{j+1}-z)\nonumber \\
 & +\tilde{r}_{N}\Theta(z-z_{N})],\nonumber \\
c_{gL}\left(z\right)= & e^{-i(k-k_{0})z}[\tilde{t}_{1}\Theta(z_{1}-z)\nonumber \\
 & +\stackrel[j=2]{N}{\sum}\tilde{t}_{j}\Theta(z-z_{j-1})\Theta(z_{j}-z)\nonumber \\
 & +\Theta\left(z-z_{N}\right)],\label{eq:RightAmplitude}
\end{align}
with $\tilde{t}_{j}$, $\tilde{r}_{j}$ being the corresponding transmission
and reflection coefficients as in the right-incident case. Substituting
Eqs. (\ref{eq:LeftAmplitude}) and (\ref{eq:RightAmplitude}) into
(\ref{eq:Schrodinger}), one readily obtains the transmission and
reflection coefficients ($t_{N}$ and $r_{1}$) for a left-incident
photon, and the transmission and reflection coefficients ($\tilde{t}_{1}$
and $\tilde{r}_{N}$) for a right-incident photon. Furthermore, the
transmission and reflection probabilities are given by $\mathcal{T}_{L}=\left|t_{N}\right|^{2}$
($\mathcal{T}_{R}=\left|\tilde{t}_{1}\right|^{2}$) and $\mathcal{R}_{L}=\left|r_{1}\right|^{2}$
($\mathcal{R}_{R}=\left|\tilde{r}_{N}\right|^{2}$) for a left-incident
(right-incident) photon, respectively.

\section*{Appendix B Lamb shift and effective decay rates for the UUEC and
BUEC regimes}

For $N>2$, we consider the simplified model where the coupling strengths
at the $i$th coupling point $g_{Li}$ ($g_{Ri}$) for the left-propagating
(right-propagating) photons are uniquely different from that {[}assumed
to be identical to $g_{L}$ ($g_{R}$){]} of all the other ($N-1$)
coupling points. A brief calculation shows that the Lamb shift is
given by 
\begin{align}
\Delta_{ls}= & \frac{g_{L}^{2}+g_{R}^{2}}{2}\left(NS_{1}-S_{N}\right)-\frac{g_{L}(g_{L}-g_{Li})+g_{R}(g_{R}-g_{Ri})}{2}\nonumber \\
 & \text{\ensuremath{\times}}\left(2S_{1}+S_{i-1}-S_{i}+S_{N-i}-S_{N-i+1}\right),\label{eq:DeltaXY_N}
\end{align}
with $S_{j}(\tilde{\phi}_{12})=\text{sin}(j\tilde{\phi}_{12})/(1-\text{cos}\tilde{\phi}_{12})$,
$j\in\mathbf{N}$, and the effective decay rates are 
\begin{align}
\Gamma_{L}\pm\Gamma_{R}= & \frac{g_{L}^{2}\pm g_{R}^{2}}{2}C_{N}+\frac{\left(g_{L}-g_{Li}\right)^{2}\pm\left(g_{R}-g_{Ri}\right)^{2}}{2}\nonumber \\
 & +\frac{g_{L}\left(g_{L}-g_{Li}\right)\pm g_{R}\left(g_{R}-g_{Ri}\right)}{2}\nonumber \\
 & \times\left(C_{i-1}-C_{i}+C_{N-i}-C_{N-i+1}\right),\label{eq:GammaXY_N}
\end{align}
with $C_{j}(\tilde{\phi}_{12})=\text{sin}^{2}\left(\frac{1}{2}j\tilde{\phi}_{12}\right)/\text{sin}^{2}\left(\frac{1}{2}\tilde{\phi}_{12}\right)$.
(1) The Lamb shifts vanish for $\text{mod}(\tilde{\phi}_{12},2\pi)=0$
due to $S_{j}(\tilde{\phi}_{12})=0$, but the decay rates in Eq. (\ref{eq:GammaXY_N})
become dependent on the number of the coupling points according to
\begin{equation}
\Gamma_{L}\pm\Gamma_{R}=\frac{\left[\left(N-1\right)g_{L}+g_{Li}\right]^{2}\pm\left[\left(N-1\right)g_{R}+g_{Ri}\right]^{2}}{2}.
\end{equation}
It follows that the transmission probabilities are\begin{widetext}
\begin{equation}
\mathcal{T}_{L(R)}=\frac{4\Delta^{2}+\left\{ \left[\left(N-1\right)g_{L}+g_{Li}\right]^{2}-\left[\left(N-1\right)g_{R}+g_{Ri}\right]^{2}\right\} ^{2}}{4\Delta^{2}+\left\{ \left[\left(N-1\right)g_{L}+g_{Li}\right]^{2}+\left[\left(N-1\right)g_{R}+g_{Ri}\right]^{2}\right\} ^{2}}.
\end{equation}
\end{widetext}(2) For $\text{mod}(\tilde{\phi}_{12},2\pi)=2m^{\text{\ensuremath{\prime}}}\pi/N$,
there exist the mathematical identities $C_{j}(\tilde{\phi}_{12})=C_{N-j}(\tilde{\phi}_{12})$
and $S_{j}\left(\tilde{\phi}_{12}\right)=-S_{N-j}\left(\tilde{\phi}_{12}\right)$
($j=0,1,...,N$), from which we readily find the Lamb shifts being
nonvanishing and $N$-dependent 
\begin{align}
\Delta_{ls}= & NS_{1}\frac{g_{L}^{2}+g_{R}^{2}}{2}-\left(S_{1}+S_{i-1}-S_{i}\right)\nonumber \\
 & \times\left[g_{L}\left(g_{L}-g_{Li}\right)+g_{R}\left(g_{R}-g_{Ri}\right)\right],\label{eq:deltaXY-1}
\end{align}
and more interestingly, the\textit{ $N$-independent} effective decay
rates 
\begin{equation}
\Gamma_{L}\pm\Gamma_{R}=\frac{1}{2}\left[(g_{L}-g_{Li})^{2}\pm(g_{R}-g_{Ri})^{2}\right].\label{eq:GammaXY_2mpi/N-1}
\end{equation}

\bibliographystyle{apsrev4-2}
\bibliography{ref}

\end{document}